\documentclass[10pt]{iopart}

\usepackage{cite} % to collapse lists of references

\usepackage{bm} % for bold math

\usepackage{color}

\newcommand{\Begeq}{\begin{equation}}
\newcommand{\Endeq}{\end{equation}}
\newcommand{\Begeqa}{\begin{eqnarray}}
\newcommand{\Endeqa}{\end{eqnarray}}

\newcommand{\mybrack}[1]{\left\{ #1 \right\}}

\renewcommand{\vec}{\bi}

\def\onedot{$\mathsurround0pt\ldotp$}
\def\cdddot{% three dots 
  \mathbin{\vcenter{\baselineskip.67ex
    \hbox{\onedot}\hbox{\onedot}\hbox{\onedot}%
  }}%
}

\newcommand{\mrm}{\mathrm}

\newcommand{\cs}{c_\mathrm{s}}
\newcommand{\pfc}{\mathcal{Z}_\mathrm{c}}
\newcommand*{\kk}{\vec{k}}
\newcommand*{\rr}{\vec{r}}
\newcommand*{\qq}{\vec{q}}
\newcommand*{\pp}{\vec{p}}
\newcommand*{\jj}{\vec{j}}
\newcommand*{\vv}{\vec{v}}

\newcommand*{\sol}{^{(\mathrm{s})}}
\newcommand*{\dmb}{^{(\mathrm{d})}}
\newcommand*{\rel}{^{(\mathrm{r})}}

\newcommand*{\nt}{^{(\mathrm{int})}}
\newcommand*{\red}{^{(\mathrm{red})}}

\newcommand*{\abgd}{{\alpha\beta\gamma\delta}}
\newcommand*{\ab}{{\alpha\beta}}

\newcommand*{\ie}{i.e. }
\newcommand*{\eg}{e.g. } 
\newcommand*{\cf}{cf. }

\def\softd{{\leavevmode\setbox1=\hbox{d}%
          \hbox to 1.05\wd1{d\kern-0.4ex{\char039}\hss}}}%cstocs

\newcommand{\MariasFirstName}{M\' aria}
\newcommand{\MariasLastName}{Luk\'a\v{c}ov\'a-Medvi\softd ov\'a}

\begin{document}

\title{﻿Systematic derivation of hydrodynamic equations
  for viscoelastic phase separation}

\author{Dominic Spiller$^1$,
  Aaron Brunk$^2$,
  Oliver Habrich$^3$,
  Herbert Egger$^3$,
  {\MariasFirstName} {\MariasLastName}$^2$
  and
  Burkhard D\"{u}nweg$^{1,4}$\footnote{Author to whom any
    correspondence should be addressed.}
  }

\address{$^1$ Max Planck Institute for Polymer Research,
  Ackermannweg 10, 55128 Mainz, Germany}

\address{$^2$ Institute of Mathematics, Johannes Gutenberg
  University Mainz, Staudingerweg 9, 55128 Mainz, Germany}

\address{$^3$ Department of Mathematics, Technical University Darmstadt, 
  Dolivostra{\ss}e 15, 64293 Darmstadt, Germany}

\address{$^4$ Department of Chemical Engineering, Monash University,
  Clayton, Victoria 3800, Australia}

\eads{
  \mailto{spiller@mpip-mainz.mpg.de},
  \mailto{abrunk@uni-mainz.de},
  \mailto{habrich@mathematik.tu-darmstadt.de},
  \mailto{egger@mathematik.tu-darmstadt.de},
  \mailto{lukacova@mathematik.uni-mainz.de},
  \mailto{duenweg@mpip-mainz.mpg.de}
}

\begin{abstract}

  We present a detailed derivation of a simple hydrodynamic two-fluid
  model, which aims at the description of the phase separation of
  non-entangled polymer solutions, where viscoelastic effects play a
  role. It is directly based upon the coarse-graining of a
  well-defined molecular model, such that all degrees of freedom have
  a clear and unambiguous molecular interpretation. The considerations
  are based upon a free-energy functional, and the dynamics is split
  into a conservative and a dissipative part, where the latter
  satisfies the Onsager relations and the Second Law of
  thermodynamics. The model is therefore fully consistent with both
  equilibrium and non-equilibrium thermodynamics. The derivation
  proceeds in two steps: Firstly, we derive an extended model
  comprising two scalar and four vector fields, such that inertial
  dynamics of the macromolecules and of the relative motion of the two
  fluids is taken into account. In the second step, we eliminate these
  inertial contributions and, as a replacement, introduce
  phenomenological dissipative terms, which can be modeled easily by
  taking into account the principles of non-equilibrium
  thermodynamics. The final simplified model comprises the momentum
  conservation equation, which includes both interfacial and elastic
  stresses, a convection-diffusion equation where interfacial and
  elastic contributions occur as well, and a suitably convected
  relaxation equation for the end-to-end vector field. In contrast to
  the traditional two-scale description that is used to derive
  rheological equations of motion, we here treat the hydrodynamic and
  the macromolecular degrees of freedom on the same
  basis. Nevertheless, the resulting model is fairly similar, though
  not fully identical, to models that have been discussed previously.
  Notably, we find a rheological constitutive equation that differs
  from the standard Oldroyd-B model. Within the framework of kinetic
  theory, this difference may be traced back to a different underlying
  statistical-mechanical ensemble that is used for averaging the
  stress. To what extent the model is able to reproduce the full
  phenomenology of viscoelastic phase separation is presently an open
  question, which shall be investigated in the future.
  
\end{abstract}


\submitto{\JPCM}

\maketitle

% two-column layout
\ioptwocol

\section{Introduction}
\label{sec:intro}

\subsection{General background}
\label{sec:GeneralBackground}

The kinetics of first-order phase transitions is a fascinating subject
of non-equilibrium thermodynamics, which has found considerable interest
in the last decades~\cite{onukiPhaseTransitionDynamics2002a}. Simple
situations are meanwhile well understood in terms of ``model A'',
``model B'', ``model H'',
etc.~\cite{brayTheoryPhaseorderingKinetics2002,
  chaikinPrinciplesCondensedMatter2000}. We are here interested in a
generalization of ``model H'', where the order parameter is conserved
and is, beyond diffusion, convectively coupled to hydrodynamic flow,
while acting back on that flow via bulk and interfacial stresses. The
prototypical physical system that is described by these equations is
an unmixing binary fluid.

The situation becomes much more involved if one of the unmixing
species is macromolecular, such that the time scale of molecular
relaxation becomes comparable to that of domain coarsening. The
interplay of the intramolecular degrees of freedom with the
macroscopic unmixing results in a plethora of interesting and
non-trivial phenomena, which were first discovered by
H. Tanaka~\cite{tanakaUnusualPhaseSeparation1993}, given the name
``viscoelastic phase separation'', and then extensively studied by him
and his collaborators~\cite{tanakaUniversalityViscoelasticPhase1996,
  tanakaViscoelasticModelPhase1997, tanakaRolesBulkRelaxation1997,
  tanakaViscoelasticModelPhase1999, tanakaViscoelasticModelPhase2000,
  tanakaViscoelasticPhaseSeparation2000a,
  nakazawaPhaseSeparationGelation2001,
  tanakaNetworkFormationViscoelastic2002,
  tanakaUniversalityViscoelasticPhase2005,
  tatenoPowerlawCoarseningNetworkforming2021};
Ref.~\cite{tanakaViscoelasticPhaseSeparation2000a} provides a review.
The crucial aspect seems to be \emph{dynamical asymmetry}, meaning
that the structural relaxation of one component is much slower than
that of the other one. Indeed, a very recent numerical investigation
by Tateno and Tanaka~\cite{tatenoPowerlawCoarseningNetworkforming2021}
reveals the phenomenon for a phase-separating colloidal
dispersion. For macromolecular systems this implies that they do not
necessarily have to be entangled. On the other hand, though,
non-entangled systems tend to exhibit only small to moderate dynamic
asymmetry, such that they show the typical phenomena of viscoelastic
phase separation only to a small (or even unobservable) degree. At any
rate, the reader should be aware that, for the purposes of the present
paper, we mean by the term ``viscoelastic phase separation'' not
necessarily the set of characteristic observations described in \eg
Ref. \cite{tanakaViscoelasticPhaseSeparation2000a}, but rather simply
phase separation in a system where coupling to viscoelasticity plays a
role.

The attempts to describe the phenomena theoretically by suitably
combining ``model H'' with viscoelastic models such as the upper
convected Maxwell model, or
Oldroyd-B~\cite{birdDynamicsPolymericLiquids1987,
  birdDynamicsPolymericLiquids1987a} are nearly as old as their
experimental investigation. Such models, which we will call here
``viscoelastic model H'' (VEMH) systems, have been considered in
nearly all of the papers cited
above~\cite{tanakaUnusualPhaseSeparation1993,
  tanakaViscoelasticModelPhase1997, tanakaRolesBulkRelaxation1997,
  tanakaViscoelasticModelPhase1999, tanakaViscoelasticModelPhase2000,
  tanakaViscoelasticPhaseSeparation2000a,
  nakazawaPhaseSeparationGelation2001,
  tanakaNetworkFormationViscoelastic2002} as well as in Refs.~\cite{
  taniguchiNetworkDomainStructure1996,
  zhouModifiedModelsPolymer2006a}, mainly building on older work by
Doi and Onuki~\cite{doiDynamicCouplingStress1992} and
Milner~\cite{milnerDynamicalTheoryConcentration1993}. While these
studies have undoubtedly advanced our understanding significantly, we
nevertheless believe that not all problems have fully been solved
already. For this reason, we wish to contribute to the field by
developing our own VEMH, and this is the main purpose of the present
paper. As in most of the existing models, we will base our reasoning
on a so-called ``two-fluid model'', where each of the components is
assigned its own velocity flow field. The central guiding principle is
that the formulation should permit a direct comparison with
microscopic or mesoscopic computer simulations, where the polymer
conformations as well as the polymer and solvent velocities appear
explicitly. In other words, it should be possible to ``measure'' each
and every dynamical variable that appears in the theory.

Before we embark on the details of the derivation, we first want to
explain why we believe that a new and fresh look at the problem may be
helpful. We therefore first outline the problems that, in our opinion,
still affect the existing literature. On the one hand, we believe that
the existing VEMHs are in some aspects a bit unsatisfactory in terms
of their physical interpretation, and we will explain these aspects in
Subsection~\ref{sec:PhysicalProblems}. On the other hand, we believe
that the existing standard methods to derive rheological constitutive
equations have a very subtle conceptual problem, and we will try to
elucidate this in Subsection~\ref{sec:RheologyConceptualProblems}. Our
work attempts to address both types of difficulty. We therefore
present not only a new set of VEMH equations, which turns out to be
fairly simple, and quite similar to existing sets, but also a somewhat
unconventional method of derivation, which, to the best of our
knowledge, has not yet been applied to the VEMH problem, and which may
turn out to be useful even beyond the VEMH context, \ie for rheology
in general. The general ``philosophy'' of our strategy will therefore
be outlined in Subsection~\ref{sec:ModelPhilosophy}. Subsection
\ref{sec:IntroOverview} will then provide an overview over the
remainder of the paper.

\subsection{Physical problems of existing VEMHs}
\label{sec:PhysicalProblems}

In Ref.~\cite{zhouModifiedModelsPolymer2006a}, which, to our
knowledge, is the most recent existing VEMH model, Zhou et al. start
from a predecessor model, due to Tanaka and
coworkers~\cite{nakazawaPhaseSeparationGelation2001} and criticize it
as violating the Second Law. They then modify the set of equations in
such a way that they are strictly and provenly dissipative. In the
typical isothermal setting of theories of phase separation, this means
that the associated free energy functional decreases monotonously with
time. This is of course a big advantage, and a property that a
physically sound model should certainly satisfy. However, another
condition for soundness is that all terms in the equations should have
clear and well-defined molecular counterparts, and that all
contributions in the free energy should scale properly, as they are
known from mechanics, thermodynamics, and polymer physics. We believe
that the model of Ref.~\cite{zhouModifiedModelsPolymer2006a} fails in
that respect. The functional features a putative ``bulk stress'' and a
putative ``shear stress'', where the ``shear stress'' enters
\emph{linearly} but the ``bulk stress'' \emph{quadratically}. As
stresses are essentially the same objects, regardless of whether they
refer to volume or to shear deformation, the powers should rather be
identical. Furthermore, it is not clear how these objects are related
to the conformations of the macromolecules. From polymer physics,
i.e., more precisely, the concept of entropic elasticity, it is known
that, for small deformations and in the absence of excluded-volume
interactions and entanglements, the elastic energy should scale
quadratically with the molecular extension, or linearly with the
conformation tensor, which is the tensor product of the molecule's
end-to-end vector with itself.

An even more subtle problem occurs in the model by Taniguchi and
Onuki~\cite{taniguchiNetworkDomainStructure1996}. Here the elastic
energy is based upon a tensor $\boldsymbol{W}$, whose meaning is
explained in the paper by
Milner~\cite{milnerDynamicalTheoryConcentration1993} as a
\emph{strain}. The idea is to start from a set of phantom Gaussian
chains, originally in equilibrium, which is being subjected to an
affine deformation that may be parameterized by $\boldsymbol{W}$.  The
partition function before and after the deformation may be calculated,
which in turn permits the calculation of the free energy as function
of $\boldsymbol{W}$. The harmonic approximation to this expression is
the elastic free energy appearing in
Ref.~\cite{taniguchiNetworkDomainStructure1996}. The problem about
this is that \emph{$\boldsymbol{W}$ is not an observable}, meaning
that its value cannot be straightforwardly inferred from the
configurations of polymer chains in, say, a Molecular Dynamics
simulation. For a solid, this would be possible, since in this case
undeformed reference positions of all the atoms are known, such that
the strain can be measured by comparing the actual atom positions with
the reference positions. In the present situation, however, we have a
fluid, such that a reference configuration simply does not exist. We
speculate that it might perhaps be possible to assign a well-defined
thermodynamic meaning to $\boldsymbol{W}$ in analogy to, say, a
chemical potential, and to construct advanced sampling schemes to
estimate its value; however, at present this must be viewed as a
completely unsolved problem. For the purposes of the present paper, we
can therefore conclude that the formulation in terms of
$\boldsymbol{W}$ is not very suitable for a direct comparison with
miscroscopic or mesoscopic simulations.

Finally, the work by Elafif et
al.~\cite{elafifRheologyDiffusionSimple1999} only discusses the
overall structure of the theory but does not specify an explicit form
of the elastic energy.

\subsection{Rheological modelling}
\label{sec:RheologyConceptualProblems}

As far as we know, there exist two major streams of research in the
goal to construct rheological models or constitutive equations for
complex fluids. These are nicely separated in terms of the two volumes
of the monograph by Bird
et~al.~\cite{birdDynamicsPolymericLiquids1987,
  birdDynamicsPolymericLiquids1987a}. On the one hand there is the
``phenomenological paradigm'' (volume 1). Here one starts from known
conservation laws and symmetry principles, which turn out to
significantly restrict the form of the equation of motion for the
stress. Nevertheless, a significant freedom remains, and this is used
to postulate simple assumptions (like linear relaxation) or to
reproduce experiments. Here one either deliberately refrains from
attempting a molecular interpretation, or invokes molecular kinetic
theory, which is however the subject of volume 2. Interestingly, such
ambiguities, which result in a certain freedom of choice, occur not
only in the construction of rheological constitutive equations but
also in the definition of two-fluid
models~\cite{pleinerGeneralNonlinear2Fluid2004}. In both cases, the
ambiguity occurs only on the purely phenomenological level, while it
is removed as soon as a well-defined molecular picture is invoked. For
rheology, this is known from kinetic
theory~\cite{birdDynamicsPolymericLiquids1987a}, while for the
two-fluid model case the same will become apparent in the course of
this paper.

The kinetic-theory approach to rheology (volume 2) may be called the
``distribution function paradigm''. Here one considers the Brownian
dynamics of model chains (\eg harmonic or FENE dumbbells, Gaussian
chains, etc.) in an external flow, within the framework of a
Fokker-Planck equation, whose solution is a time-dependent
distribution function in conformation space. This function may then be
used for defining time-dependent thermal averages of observables like
the conformation tensor or the virial. The resulting expression of the
stress is then fed into the momentum equation, while the Fokker-Planck
equation results in relaxation equations for observables (\eg the
conformation tensor components) that are needed to obtain a closed
set. By construction, the method does provide a clear molecular
interpretation.

The conceptual problem that we see in that approach lies in the fact
that the statistical-mechanical averages are done without any
\emph{constraint}. To elucidate what we mean by this point, let us,
for a moment, go back to the simple hydrodynamics of a monatomic
fluid, and its root in statistical mechanics. The idea is that we may
consider a small volume element that, despite its smallness, contains
a large number of particles such that we may reasonably treat it in
terms of statistical physics. Furthermore, hydrodynamics assumes that,
within the volume element, all variables are in \emph{local
  equilibrium}, \emph{except} the hydrodynamic variables like the mass
(or particle number), the total energy, and the total momentum. These
latter variables may therefore be viewed as \emph{constraints that
  define the appropriate thermodynamic ensemble} for the local volume
element. Now, if we introduce, say, the chain conformation tensor as
an additional variable into the hydrodynamic description, then, in our
opinion, this variable should occur as a corresponding additional
constraint in the definition of the thermodynamic ensemble.  However,
here one should watch out that only \emph{independent} variables are
used as constraints. In other words: the stress should not be
calculated as a full average over all chain conformations, but rather
as a restricted average over the constrained ensemble. In this
context, note that in standard statistical physics one typically
ignores this problem by relying on the equivalence of constrained and
non-constrained ensembles. However, this equivalence holds only in the
asymptotic limit of infinite system size, with error terms which scale
linearly with the inverse number of the involved degrees of
freedom. To extend this notion down to the scale of a single
macromolecule seems somewhat problematic, and we believe that this is
more than a petty formality. Indeed, the different ensembles do result
in different equations of motion on the macroscale. In more explicit
terms, this difference is worked out in \ref{sec:EnsembleProblems}. As
one sees there, the difference boils down, for a Hookean dumbbell, to
the difference between \emph{mechanical} and \emph{thermal}
equilibrium, which is substantial for a strongly fluctuating
soft-matter system like a polymer solution. We believe that a
consistent way to treat the effects of fluctuations is \emph{not} by
doing an unconstrained average (as in the derivation of the standard
Oldroyd-B model) but rather by augmenting \emph{all} macroscopic
equations with Langevin noise. As we will see below, the equations of
motion that result from the constrained average (see
\ref{sec:EnsembleProblems}) are \emph{identical} to those that are
derived in the present paper, using a completely different approach
--- but different from the standard Oldroyd-B model. In the present
paper, we ignore Langevin noise in the macroscopic equations and defer
that aspect to future work.

\subsection{Basic ``philosophy'' of our model}
\label{sec:ModelPhilosophy}

The route that we take is somewhat different from the usual
rheological literature, and it may be called a ``coarse-graining
paradigm''. Its spirit is similar (although technically simpler) to
analogous developments in the theory of liquid
crystals~\cite{starkPoissonbracketApproachDynamics2003}. We start from
a microscopic model and then define microscopic expressions for
various fields. To give an example for illustration: For a monatomic
fluid of particles located at positions $\vec{r}_i$, the expression
for the particle number density at position $\vec{r}$ would be $\sum_i
\delta (\vec{r} - \vec{r}_i)$. We then subject that system to
coarse-graining and derive corresponding field-theoretic equations of
motion. This is (roughly spoken --- for details see below) done by (i)
mapping the microscopic Hamiltonian onto the corresponding
field-theoretic one; (ii) doing an analogous mapping from a
microscopic dissipation rate to a field-theoretic one; (iii)
postulating a dissipative Hamiltonian system; (iv) finding the
Hamiltonian part of the dynamics via the Poisson brackets of the
fields, which in turn are found via reference to the microscopic
counterparts; and (v) reading off the dissipative part of the dynamics
directly from the dissipation rate. After that, several fields are
identified as fast variables and therefore eliminated adiabatically,
where unknown terms are replaced by phenomenological dissipative
terms. At the end, we arrive at a model that looks similar to the
existing VEMH equations, but differs in various subtle aspects.

We believe that this is a quite powerful approach and has potential
beyond the immediate application to the VEMH problem. The Poisson
bracket formalism has proven extremely fruitful in
hydrodynamics~\cite{ salmonHamiltonianFluidMechanics1988,
  zakharovHamiltonianFormalismNonlinear1997a,
  morrisonHamiltonianDescriptionIdeal1998,
  berisPoissonBracketFormulation1990,
  berisPoissonBracketFormulation1990a,
  edwardsNoncanonicalPoissonBracket1991,
  berisThermodynamicsFlowingSystems1994}; readers not familiar with
that formalism are advised to briefly consult
\ref{sec:poisson_general} and \ref{sec:poisson_euler}. Due to the
construction via a dissipative Hamiltonian system, whose equation of
motion can be found as a by-product of the GENERIC
formalism~\cite{ottingerEquilibriumThermodynamics2005,
  grmelaDynamicsThermodynamicsComplex1997,
  ottingerDynamicsThermodynamicsComplex1997, grmelaWhyGENERIC2010}, we
automatically make sure that the Second Law is satisfied. Fully in
line with the arguments put forward in
Ref.~\cite{zhouModifiedModelsPolymer2006a}, we consider this as very
important both from the point of view of fundamental physical
consistency, but also from the point of view of mathematical analysis
and derivation of stable and convergent numerical
algorithms~\cite{lukacova-medvidovaEnergyStableNumerical2016,
  strasserEnergystableLinear2019}.

The hydrodynamic model that we construct in the first step (before
the adiabatic elimination of the fast variables) comprises two scalar
and four vector fields, which are chosen in order to be able to take
full advantage of Hamiltonian dynamics and the Poisson bracket
formalism. The fields are (i) the mass densities of the two components
(solvent and polymer), (ii) the two associated momentum densities
(note we study a two-fluid model), (iii) a vector field of molecular
end-to-end vectors, and (iv) an associated momentum density. The fast
variables that we eliminate in the second step are the relative
velocity, the internal molecular momentum, plus the fluctuations of
the total density (and the associated longitudinal modes).

The elastic Hamiltonian is quadratic, but not in terms of the
conformation tensor but rather the end-to-end vector, which we use as
the elementary field to describe the internal relaxation of the
macromolecules. The equation of motion for the end-to-end vector can,
in the final stage, be transformed to an equation for the conformation
tensor. Except for constant prefactors, the stress tensor is just the
product of the conformation tensor and the polymer density (which
makes perfect sense, since there should be no elastic stress in
spatial regions where there is no polymer). All in all, our new model
has the big advantage that (i) all degrees of freedom have a
well-defined molecular meaning, and (ii) its derivation has a solid
foundation in Hamiltonian dynamics and non-equilibrium
thermodynamics. As a limitation, the reader should however notice that
entanglements are not considered by the present model.

Phenomenological closures are imposed only in the second step, where
the fast variables are removed, while in the first step (the
derivation of the six-field hydrodynamics) no such closure
occurs. This is possible since our microscopic model is significantly
simplified, compared to a truly atomistic system. In essence, it is
our computer model for polymer-solvent
systems~\cite{dunwegLatticeBoltzmann2009,
  ahlrichsSimulationSingle1999, tretyakovImprovedDissipative2017}, in
which bead-spring polymer chains are simulated by Molecular Dynamics
(MD), while the solvent is represented by an ideal-gas type
hydrodynamic background, simulated by Lattice Boltzmann (LB). The
coupling between MD and LB is of a purely dissipative nature,
facilitated by assigning a Stokes friction coefficient to the
beads. Starting from there, we simplify the model even further by
replacing the polymer chains by phantom Hookean dumbbells with
non-bonded interactions being replaced by a Van der Waals Mean Field
model (see below). This model is so simple that the construction of
the six-field hydrodynamics can be done exactly, except, of course,
for the inevitable approximations that come from the field-theoretic
``smoothing'' of the microscopic fields.

\subsection{Outline of the remainder of the paper}
\label{sec:IntroOverview}

Section~\ref{sec:model} defines the microscopic model and, from there,
the fields and the field-theoretic
Hamiltonian. Section~\ref{sec:GENERIC} discusses the basic equation of
motion of isothermal dissipative Hamiltonian systems, which is then
used to derive the reversible (Section~\ref{sec:hamil}) and the
irreversible (Section~\ref{sec:dissip}) parts of the six-field
hydrodynamics, which is summarized in Section~\ref{sec:FinalSet}. In
Section~\ref{sec:transformation} we transform to new variables (total
density and density contrast, barycentric velocity and relative
velocity). This is the basis for the application of various
approximations and simplifications (outlined in
Section~\ref{sec:approx}), which, in essence, boil down to adiabatic
elimination of fast variables. Section~\ref{sec:simplified} then
provides a detailed analysis of the resulting VEMH system, showing
that it is compatible with the Second Law and the Onsager symmetry
relations. It is also in this section where the phenomenological terms
are specified; again the Second Law is here the essential guiding
principle. Section~\ref{sec:vanderwaals} sketches the polymer version
of Van der Waals theory, upon which the bulk part of the Hamiltonian
is based, and Section~\ref{sec:conclus} provides a few concluding
remarks. \ref{sec:poisson_general} and \ref{sec:poisson_euler} are
meant as background material, where the Poisson bracket formalism, and
its application to derive the standard Euler equations, are
outlined. Finally, \ref{sec:EnsembleProblems} further elucidates the
difference between constrained and unconstrained averages in the
derivation of rheological equations.

At this point, let us emphasize that the present paper is concerned
with the derivation of rheological equations of motion, and \emph{not}
with the question if (or to what extent) the resulting model is able
to reproduce the existing experimental observations that are known for
viscoelastic phase separation. In our opinion, this question can only
be answered by detailed computer simulations, which are however beyond
the scope of the present investigation.

\section{Model}
\label{sec:model}

\subsection{Solvent}

The dynamics of the solvent degrees of freedom is always (\ie  both
on the microscopic level and on the fully field-theoretic level)
represented by the isothermal Navier-Stokes equations
\Begeq
\label{eq:MassConservationSolvent}
\partial_t \rho\sol + \nabla \cdot \vec{j}\sol = 0 ,
\Endeq
\Begeqa
\label{eq:NavierStokesSolventI}
&&
\partial_t j_\alpha\sol +
\partial_\beta \left( \rho\sol v_\alpha\sol v_\beta\sol \right)
\\
\nonumber
& = & - \partial_\alpha p\sol
+ \eta_\abgd \partial_\beta \partial_\gamma v\sol_\delta + f_\alpha.
\Endeqa
Here, the upper index ``(s)'' refers to the solvent, such that
$\rho\sol$ is the solvent mass density etc.. The symbol $\partial_t$
denotes the time derivative $\partial / \partial t$, and similarly
$\partial_\alpha$ the spatial derivative $\partial / \partial
r_\alpha$. Greek subscripts are Cartesian indexes for which we assume
the Einstein summation convention. $\vec{j}\sol$ is the solvent
momentum density, related to the solvent velocity flow field
$\vec{v}\sol$ via $\vec{j}\sol = \rho\sol \vec{v}\sol$. $p\sol$ is the
solvent partial pressure, for which the LB model prescribes the
equation of state of an ideal gas, $p\sol = \rho\sol\cs^2$, where
$\cs$ is the speed of sound of the LB fluid. $\eta_\abgd$ is the
fourth-rank viscosity tensor, which for an isotropic Newtonian fluid
(like our LB fluid) reads
\Begeq
\eta_\abgd =
\left( \eta_{\mathrm{V}} - \frac{2}{3} \eta_{\mathrm{s}} \right)
\delta_{\alpha \beta} \delta_{\gamma \delta} +
\eta_{\mathrm{s}} \left( \delta_{\alpha \gamma} \delta_{\beta \delta}
+ \delta_{\alpha \delta} \delta_{\beta \gamma} \right)
\Endeq
with bulk viscosity $\eta_{\mathrm{V}}$ and shear viscosity
$\eta_{\mathrm{s}}$. Finally, $\vec{f}$ is a dissipative force
density, which comes from the coupling to the polymer component. If
the latter is being described in terms of Stokes beads, as on the
microscopic level, $\vec{f}$ can be written down explicitly. On the
fully field-theoretic level, the corresponding expression must be
constructed with care; Section~\ref{sec:dissip} will outline how to do
that.

In a shorthand notation, we may write
Eq.~\ref{eq:NavierStokesSolventI} as
\Begeqa
&&
\partial_t \vec{j}\sol + \nabla \cdot \left(
\rho\sol \vec{v}\sol \vec{v}\sol \right)
\\
\nonumber
& = &
- \nabla p\sol
+ \boldsymbol{\eta} \cdddot \nabla \nabla \vec{v}\sol
+ \vec{f} .
\Endeqa

The Hamiltonian of the solvent is
\Begeq
H\sol =
\int d^3 \vec{r} \left( \frac{\left( \vec{j}\sol \right)^2}{2 \rho\sol}
+ f\sol \right) ,
\Endeq
where $f\sol$ is the free energy density of the solvent; note that in
our isothermal setting we need to interpret the Hamiltonian as the
Helmholtz free energy. The pressure is derived from the free energy
via
\Begeqa
p\sol & = &
\left( \rho\sol \right)^2 \frac{\partial}{\partial \rho\sol}
\left( \frac{f\sol}{\rho\sol} \right) ,
\\
\nabla p\sol & = &
\rho\sol \nabla \left( \frac{\partial f\sol}{\partial \rho\sol} \right) .
\Endeqa

\subsection{Polymer component: Microscopic description}

Each polymer chain is represented by a Hookean dumbbell. The
center-of-mass coordinate of dumbbell number $i$ is denoted by
$\vec{r}_i\dmb$, and the two beads are located at the positions
$\vec{r}_i\dmb \pm \vec{q}_i / 2$, such that $\vec{q}_i$ is the
connector vector of the two beads. If we assign a mass $m$ to each
bead, then the total mass is $m\dmb = 2 m$ and the reduced mass $m\rel
= m / 2$. We denote $\Gamma = m\rel / m\dmb$, which takes the value
$\Gamma = 1/4$ for our model; however, we would like to keep $\Gamma$
as a parameter that may perhaps be adjusted. Momenta associated with
the dumbbell are $\vec{p}_i\dmb = m\dmb \dot{\vec{r}}_i\dmb$ for the
center-of-mass motion and $\vec{p}_i\rel = m\rel \dot{\vec{q}}_i =
\Gamma m\dmb \dot{\vec{q}}_i$ for the relative motion. Furthermore,
the spring constant of the dumbbells is denoted by $k$. The
Hamiltonian of the dumbbell system is thus given by
\Begeqa
\nonumber
\hat{H}\dmb & = &
\sum_i \left( \frac{\left( \vec{p}_i\dmb \right)^2}{2 m\dmb}
+ \Gamma^{-1} \frac{\left( \vec{p}_i\rel \right)^2}{2 m\dmb}
+ \frac{k}{2} \vec{q}_i^2 \right)
\\
& + & U_{\rm{nb}} \left( \{ \vec{r}_i\dmb \}, \{ \vec{q}_i \} \right) ,
\Endeqa
where $U_{\rm{nb}}$ is the non-bonded interaction potential, which we
do not need to specify in detail. As a typical example of what we mean
by this, the reader may assume a pairwise bead-bead interaction that
acts in the same way for all pairs of beads. The associated potential
is then characterized by a strong repulsive core at short
interparticle distances, and an attractive tail at larger
distances. We may then assume that the strength of the attractive part
parameterizes the solvent quality, such that the polymer component
falls out of solution as soon as the attraction exceeds a critical
value.

It is clear that such a model is a simple caricature of a real polymer
solution, and that it certainly is unable to describe
entanglements. Nevertheless, it is the standard starting point of many
theoretical developments of
rheology~\cite{birdDynamicsPolymericLiquids1987a}.

Since the interaction between solvent and dumbbell system is
purely dissipative, the total Hamiltonian is just the sum of
the individual Hamiltonians
\Begeq
\hat{H} = H\sol + \hat{H}\dmb .
\\
\Endeq

\subsection{Polymer component: Microscopic expressions for fields}

We can now construct microscopic expressions for various fields
associated with the dumbbell degrees of freedom. These are:
(i) the dumbbell mass density
\Begeq
  \rho\dmb (\rr) = m\dmb \sum_i \delta \left(\rr - \vec{r}_i\dmb \right) ,
\Endeq
(ii) the dumbbell momentum density
\Begeq
  \jj\dmb(\rr) = \sum_i \pp\dmb_i \delta \left(\rr - \vec{r}_i\dmb \right) ,
\Endeq
and analogous vector fields for the polymer extension, \ie (iii) the
elastic force density
\Begeq
  \kk\rel(\rr) = k \sum_i \vec{q}_i \delta \left(\rr - \vec{r}_i\dmb \right) ,
\Endeq
and (iv) the relative momentum density
\Begeq
  \jj\rel(\rr) = \sum_i \pp\rel_i \delta \left(\rr - \vec{r}_i\dmb \right) .
\Endeq

\subsection{Polymer component: Field-theoretic description}
\label{sec:DumbbellFieldTheory}

After coarse-graining (``smoothing''), these fields are replaced by
continuous functions, and we will consider these smoothed fields,
together with $\rho\sol$ and $\vec{j}\sol$, as the independent degrees
of freedom of the continuum theory that we wish to construct. We thus
assume that the orientation and stretching of the polymer chains in
the flow can be reasonably described by a smooth vector field
$\kk\rel(\rr)$, and the internal motions by a corresponding smooth
vector field $\jj\rel(\rr)$. Now, it is clear that on the microscopic
level the physics does not change if we replace $\vec{q}_i$ by $-
\vec{q}_i$ (we call this a ``flip''), which also means
$\dot{\vec{q}}_i \to - \dot{\vec{q}}_i$ and $\vec{p}_i\rel \to -
\vec{p}_i\rel$. Therefore, on the field-theoretic level, the physics
should not change under the analogous flip transformation
$\kk\rel(\rr) \to - \kk\rel(\rr)$ and $\jj\rel(\rr) \to -
\jj\rel(\rr)$. This, in turn, means that the model should be ``flip
covariant'', meaning that, within a given equation of motion, all
occuring terms must have the same transformation behavior under flip
--- they must all be even, or they must all be odd. As can be easily
checked throughout the derivation that we will present below, this is
indeed the case, step by step. It does \emph{not} imply, though, that
only even terms are permitted. To illustrate that point, just consider
Newton's equation of motion, where both the force and the acceleration
have the same transformation behavior under reflection --- but both
are odd, not even.

The intuitive picture that we associate with the assumption of a
\emph{smooth} vector field $\kk\rel(\rr)$ is this: We arbitrarily pick
one particular dumbbell, and then arbitrarily pick one of its two
possible orientations. This defines the connector vector for that
particular dumbbell. Now we go to another dumbbell in the immediate
vicinity of the first one, and again pick its orientation.  We are now
no longer free to pick it arbitrarily, but rather must choose it in
such a way that its alignment with the already assigned orientation is
as good as possible. This is dictated by the requirement that the
coarse-grained vector field is supposed to be \emph{smooth}. In this
way, we scan the whole system and assign one well-defined orientation
after the other. In some cases, frustration might occur, but we assume
that such cases are rare. The resulting connector vectors enter the
definition of the field $\kk\rel(\rr)$. We believe that, with this
picture in mind, the assumption of a smooth vector field is quite
reasonable.

At this point, one might ask if it would not be more advisable to
rather assign a conformation tensor to each molecule (this is even
under flip), and then subject these quantities to coarse-graining,
such that one obtains a smooth conformation tensor field. This would
obviously eliminate the difficulties mentioned above. However, upon
considering this idea in more detail, we found ourselves unable to
construct suitable canonical momenta, and a Hamiltonian theory based
upon them, without running into severe algebraic difficulties, of
which we do not know if they can be resolved or not. The main problem
is that the components of the conformation tensor are not all
independent; rather the very construction of the tensor implies that
it has only three independent parameters. A similar difficulty also
appears in the Fokker-Planck description, see
\ref{sec:EnsembleProblems}. We therefore found it much easier to
rather base our considerations on vector fields.

The transition from the microscopic model to the field theory then
consists of replacing the dumbbell Hamiltonian by an analogous
field-theoretic expression:
\Begeqa
H\dmb & = &
\int d^3 \vec{r}
\left[
  \frac{\left( \vec{j}\dmb \right)^2}{2 \rho\dmb}
+ \Gamma^{-1} \frac{\left( \vec{j}\rel \right)^2}{2 \rho\dmb}
\right.
\\
\nonumber
& &
\left.
+ \frac{m\dmb}{k} \frac{\left( \kk\rel \right)^2}{2 \rho\dmb}
+ f\dmb + \frac{\kappa}{2} \left( \nabla \rho\dmb \right)^2
\right] .
\Endeqa
Here the first two terms describe the kinetic energy of the
center-of-mass motion and the relative motion, respectively, while the
third term describes the elastic (or spring) energy. The fourth and
fifth term are the replacement for the non-bonded interaction
$U_{\rm(nb)}$, where $f\dmb$ is the bulk part of the configurational
free energy density. The last term is meant to penalize the occurence
of interfaces; the parameter $\kappa$ is called ``interfacial
stiffness''. Such terms always occur in the field-theoretic
description of phase separation (both statics and
dynamics)~\cite{brayTheoryPhaseorderingKinetics2002} and there is a
well-defined statistical-mechanical procedure to derive them. For a
simple approach the reader is referred to
Ref.~\cite{langerIntroductionKineticsFirstorder1992}, while a more
advanced machinery is found in
Ref.~\cite{ivanchenkoPhysicsCriticalFluctuations1995}. For the bulk
free energy density we will assume that $f\dmb = f\dmb (\rho\dmb)$ and
that $f\dmb$ can be constructed from a Van der Waals model, which is
the simplest well-known model for a fluid that undergoes a gas-liquid
transition. This choice is inspired by the fact that the solvent is
just an ideal gas (\ie thermodynamically inert), such that the
decomposition between polymer and solvent may be viewed (from the
point of view of thermodynamics, not hydrodynamics) as just a
gas-liquid transition of the polymer component.  The structure of the
theory is completely independent of the precise form of $f\dmb$;
therefore one may as well assume a different function for $f\dmb$
(possibly even an empirical function derived from simulation results).

After this ``smoothing operation'', we can then define further fields:
(i) the dumbbell velocity flow field
\Begeq
\vv\dmb(\rr) = \frac{\jj\dmb(\rr)}{\rho\dmb(\rr)} =
\frac{\delta H}{\delta \jj\dmb(\rr)} 
\Endeq
(here $\delta \ldots / \delta \ldots$ denotes the functional
derivative), (ii) the ``relative flow field''
\Begeq
\vv\rel(\rr) = \Gamma^{-1} \frac{\jj\rel(\rr)}{\rho\dmb(\rr)} =
\frac{\delta H}{\delta \jj\rel(\rr)} ,
\Endeq
and (iii) the ``extension field''
\Begeq
\vec{q}(\rr) = \frac{m\dmb}{k} \frac{\kk\rel(\rr)}{\rho\dmb(\rr)} =
\frac{\delta H}{\delta \kk\rel(\rr)} .
\Endeq
We also note
\Begeqa
\frac{\delta H}{\delta \rho\dmb}
& = &
\frac{\partial f\dmb}{\partial \rho\dmb}
- \kappa \nabla^2 \rho\dmb
\\
\nonumber
&&
- \frac{1}{2} \left[
  \left( \vv\dmb \right)^2 + \Gamma \left( \vv\rel \right)^2
  + \frac{k}{m\dmb} \qq^2 \right] .
\Endeqa

\section{General equation of motion}
\label{sec:GENERIC}

In the present paper, we will be concerned with equations of motion for
a set of fields $\Phi_i = \Phi_i (\vec{r}, t)$ in three-dimensional
space, of the form
\Begeq
\label{eq:BasicDynamics}
\partial_t \Phi_i = {\cal L}_i - {\cal M}_i .
\Endeq
We will assume a domain with periodic boundary conditions, such that
integrations by parts will never involve surface terms. The symbols
${\cal L}_i = {\cal L}_i \left( \left\{ \Phi_k \right\}, \vec{r}
\right) = {\cal L}_i (\vec{r})$, ${\cal M}_i = {\cal M}_i \left(
\left\{ \Phi_k \right\}, \vec{r} \right) = {\cal M}_i (\vec{r})$
denote fields that depend on the field degrees of freedom and describe
the dynamics. The functional $H = H \left( \left\{ \Phi_k \right\}
\right)$ is of central importance, and we will call it the Hamiltonian
of the system. Since we study the system in the isothermal ensemble,
it may also be called the (Helmholtz) free energy functional. We
require that the dynamics is dissipative, \ie that $dH / dt \le
0$. The fields ${\cal L}_i$ and ${\cal M}_i$ have been introduced to
denote the conservative (${\cal L}_i$) and the dissipative (${\cal
  M}_i$) parts of the dynamics. Obviously,
\Begeq
\frac{d H}{d t} = \sum_i \int d^3\vec{r} \, \left(
{\cal L}_i (\vec{r}) -
{\cal M}_i (\vec{r}) \right)
\frac{\delta H}{\delta \Phi_i (\vec{r})} .
\Endeq
Furthermore, we require
\Begeqa
\label{eq:GENERICConservativeCondition}
\sum_i \int d^3\vec{r} \, {\cal L}_i (\vec{r})
\frac{\delta H}{\delta \Phi_i (\vec{r})} & = & 0 ,
\\
\label{eq:GENERICDissipativeCondition}
\sum_i \int d^3\vec{r} \, {\cal M}_i (\vec{r})
\frac{\delta H}{\delta \Phi_i (\vec{r})} & \ge & 0 .
\Endeqa
We will always require that the dissipative terms can be described in
terms of linear Onsager theory, where the dissipative responses ${\cal
  M}_i$ are proportional to the driving forces $\delta H / \delta
\Phi_j$:
\Begeq
\label{eq:OnsagerForm}
{\cal M}_i (\vec{r}) = \sum_j \int d^3\vec{r}' \,M_{ij} (\vec{r}, \vec{r}') 
\frac{\delta H}{\delta \Phi_j (\vec{r}')} ,
\Endeq
where the elements $M_{ij}$ form a matrix that is symmetric with
respect to the simultaneous exchanges $i \leftrightarrow j$, $\vec{r}
\leftrightarrow \vec{r}'$; this expresses the Onsager reciprocity
relations. Furthermore, the Second Law requires that the matrix is
positive-semidefinite. The total dissipation rate therefore
takes the form
\Begeqa
&&
\label{eq:GENERICDissipationRate}
\frac{d H}{dt}
\\
\nonumber
& = & - \sum_{ij} \int d^3 \vec{r} \int d^3 \vec{r}'
\frac{\delta H}{\delta \Phi_i (\vec{r})}
M_{ij} (\vec{r}, \vec{r}')
\frac{\delta H}{\delta \Phi_j (\vec{r}')} \le 0 .
\Endeqa
If in addition the conservative part of the dynamics has a Hamiltonian
structure, then we have a dissipative Hamiltonian system. In this
case, the Poisson brackets $\mybrack{ \Phi_i (\vec{r}), \Phi_j
  (\vec{r}') }$ form a closed system (\ie they can be expressed
in terms of the fields $\Phi_i$, with no reference to additional
``hidden'' degrees of freedom), and we have
\Begeq
\label{eq:PoissonBracketForm}
{\cal L}_i (\vec{r}) = \sum_j \int d^3 \vec{r}' \,
\mybrack{ \Phi_i (\vec{r}), \Phi_j (\vec{r}') }
\frac{\delta H}{\delta \Phi_j (\vec{r}')} .
\Endeq
For more details on the Poisson bracket formalism, see
\ref{sec:poisson_general} and \ref{sec:poisson_euler}. The
conservative nature of this dynamics is then a direct consequence of
the antisymmetry of the Poisson brackets. It should be noted that
Eq.~\ref{eq:BasicDynamics} with the specific forms
Eqs.~\ref{eq:OnsagerForm} and \ref{eq:PoissonBracketForm} can be
derived from the GENERIC
formalism~\cite{ottingerEquilibriumThermodynamics2005,
  grmelaDynamicsThermodynamicsComplex1997,
  ottingerDynamicsThermodynamicsComplex1997, grmelaWhyGENERIC2010}.

We will use these general considerations in two ways. In the first
part of the paper, where we consider the six-field hydrodynamics, we
will be able to construct, by reference to the underlying microsopic
model, the Poisson brackets, the functional derivatives of the
Hamiltonian, and the dissipation rate $dH / dt$. The latter can be
written in a form that matches Eq.~\ref{eq:GENERICDissipationRate},
which will enable us to read off the matrix elements $M_{ij}$. In the
second part, \ie the VEMH where the fast degrees of freedom have
been eliminated, we know ${\cal L}_i$ (and we can explicitly show that
this is indeed conservative), but we know ${\cal M}_i$ only up to
unknown phenomenological terms. We can then find expressions
for these terms by assuming a simple but consistent model, where
the matrix is diagonal and positive-definite.

\section{Equations of motion I: Hamiltonian part}
\label{sec:hamil}

We first focus on the Hamiltonian part of the dynamics. We note that
on the Hamiltonian level our model implies that the polymer system and
the solvent system are completely decoupled, and may therefore be
treated separately. For the solvent system, the conservative part of
the dynamics is given by the Euler equations. These are known to be
Hamiltonian, and this can be shown straightforwardly by means of the
Poisson bracket formalism, see \ref{sec:poisson_euler}.

We therefore need to execute the same methodology for the dumbbell
system. The functional derivatives of the Hamiltonian have already
been evaluated in Section~\ref{sec:model}; therefore the next step
is to calculate the Poisson brackets by insertion of the microscpic
expressions, plus reference to the Poisson brackets of coordinates
and momenta, as they are known from classical mechanics. The evaluation
is somewhat tedious but straightforward and yields
\Begeqa
  \mybrack{\rho\dmb(\rr), j\dmb_\beta(\rr')}
  & = & -\rho\dmb(\rr') \partial_\beta\delta(\rr-\rr'),
  \\
  \mybrack{k\rel_\alpha(\rr), j\dmb_\beta(\rr')}
  & = & -k\rel_\alpha(\rr') \partial_\beta\delta(\rr-\rr'),
  \\
  \mybrack{j\rel_\alpha(\rr), j\dmb_\beta(\rr')}
  & = & -j\rel_\alpha(\rr') \partial_\beta\delta(\rr-\rr'),
  \\
  \mybrack{j\dmb_\alpha(\rr), j\dmb_\beta(\rr')}
  & = &
  j\dmb_\beta(\rr) \partial'_\alpha \delta(\rr-\rr')
  \\
  \nonumber
  &&
  -
  j\dmb_\alpha(\rr') \partial_\beta\delta(\rr-\rr'),
  \\
  \mybrack{k_\alpha\rel(\rr), j\rel_\beta(\rr')}
  & = & \delta_{\ab} \frac{k}{m\dmb} \rho\dmb(\rr') \delta(\rr-\rr') .
\Endeqa
The remaining Poisson brackets that have not been listed simply
vanish.

We then insert the Poisson brackets, plus the functional derivatives
of the Hamiltonian with respect to the fields, into the general
equation of motion. After some algebra (quite a few terms cancel) we
arrive at the following set.

For the dumbbell density and the dumbbell momentum density, we recover
the standard Euler equations (\cf also \ref{sec:poisson_euler}),
augmented by the interfacial force term:
\Begeq
\partial_t \rho\dmb + \nabla \cdot \vec{j}\dmb = 0 ,
\Endeq
\Begeqa
&&
\partial_t j_\alpha\dmb +
\partial_\beta \left( \rho\dmb v_\alpha\dmb v_\beta\dmb \right)
\\
\nonumber
& = & - \partial_\alpha p\dmb +
\kappa \rho\dmb \partial_\alpha \nabla^2 \rho\dmb ,
\Endeqa
where $p\dmb$ is the dumbbell partial pressure, defined as
\Begeq
p\dmb = \left( \rho\dmb \right)^2 \frac{\partial}{\partial \rho\dmb}
\left( \frac{f\dmb}{\rho\dmb} \right) .
\Endeq
It should be noted that the interfacial force term does conserve the
total momentum, \ie its spatial integral is zero, as can be shown
by integration-by-parts.

The equation of motion for the force density is found to be
\Begeq
\partial_t k_\alpha\rel + \partial_\beta \left(
k_\alpha\rel v_\beta\dmb \right) = \frac{k}{m\rel} j_\alpha\rel ;
\Endeq
this takes a somewhat more intuitive form after transforming to the
corresponding equation for the extension field, which reads, by
taking into account the mass conservation equation for $\rho\dmb$,
\Begeq
\partial_t q_\alpha + v_\beta\dmb \partial_\beta q_\alpha = v_\alpha\rel ;
\Endeq
note that the left hand side is just the convective derivative
of $q_\alpha$.

Finally, we find for the relative momentum density
\Begeq
\partial_t j_\alpha\rel + \partial_\beta \left(
j_\alpha\rel v_\beta\dmb \right) = - k_\alpha\rel .
\Endeq
Again transforming to the corresponding velocity field,
we find
\Begeq
m\rel \left( \partial_t v_\alpha\rel
+ v_\beta\dmb \partial_\beta v_\alpha\rel \right)
= - k q_\alpha ,
\Endeq
which is essentially Newton's equation of motion for the oscillator,
taking into account convection with $\vec{v}\dmb$.

In summary, we thus find a set of equations which has not only been
derived with a well-founded formalism, but is also intuitively quite
plausible.

\section{Equations of motion II: Dissipative part}
\label{sec:dissip}

To take into account dissipation, and in particular the dissipative
coupling of the dumbbell system to the background solvent fluid, we
first need to consider the dumbbell number $i$ in the solvent flow
field $\vec{v}\sol$. The two beads are located at the positions
$\vec{r}_i\dmb \pm \vec{q}_i / 2$, which means that the relevant terms
for the coupling are proportional to the difference between bead
velocity $\dot{\vec{r}}_i\dmb \pm \dot{\vec{q}}_i / 2$ and the flow
velocity at the position of the bead, $\vec{v}\sol \left(
\vec{r}_i\dmb \pm \vec{q}_i / 2 \right)$. Ignoring the (Hamiltonian)
part that comes from the spring force, we may thus write down the
equations of motion for the two beads:
\Begeqa
m \frac{d}{dt} \left( \dot{\vec{r}}_i\dmb + \dot{\vec{q}}_i / 2 \right)
& = & - \zeta \vec{u}_i^{(1)} ,
\\
m \frac{d}{dt} \left( \dot{\vec{r}}_i\dmb - \dot{\vec{q}}_i / 2 \right)
& = & - \zeta \vec{u}_i^{(2)} ;
\Endeqa
here $m$ denotes the bead mass and $\zeta$ is the friction
coefficient, whose value controls the strength of the coupling,
while
\Begeqa
\vec{u}_i^{(1)} & = & \dot{\vec{r}}_i\dmb + \dot{\vec{q}}_i / 2
  - \vec{v}\sol \left( \vec{r}_i\dmb + \vec{q}_i / 2 \right) ,
\\
\vec{u}_i^{(2)} & = & \dot{\vec{r}}_i\dmb - \dot{\vec{q}}_i / 2
  - \vec{v}\sol \left( \vec{r}_i\dmb - \vec{q}_i / 2 \right) .
\Endeqa
Introducing
\Begeqa
\vec{u}_i^{(0)}
& = &
\left( \vec{u}_i^{(1)} + \vec{u}_i^{(2)} \right) / 2
\\
\nonumber
& = &
\dot{\vec{r}}_i\dmb -
\\
\nonumber
&&
\left[
  \vec{v}\sol \left( \vec{r}_i\dmb + \vec{q}_i / 2 \right) +
  \vec{v}\sol \left( \vec{r}_i\dmb - \vec{q}_i / 2 \right)
\right] / 2 ,
\\
\Delta \vec{u}_i
& = &
\vec{u}_i^{(1)} - \vec{u}_i^{(2)}
\\
\nonumber
& = &
\dot{\vec{q}}_i -
\\
\nonumber
&&
\left[
  \vec{v}\sol \left( \vec{r}_i\dmb + \vec{q}_i / 2 \right) -
  \vec{v}\sol \left( \vec{r}_i\dmb - \vec{q}_i / 2 \right)
\right] ,
\Endeqa
and a relaxation time $\tau$ associated with the coupling, $\tau = m /
\zeta$, we may rewrite the equations as
\Begeqa
\tau \frac{d}{dt} \dot{\vec{r}}_i\dmb
& = & - \vec{u}_i^{(0)} ,
\\
\tau \frac{d}{dt} \dot{\vec{q}}_i 
& = & - \Delta \vec{u}_i .
\Endeqa
Scaling the equations with the masses $m\dmb$ and $m\rel$, respectively,
we find
\Begeqa
\tau \frac{d}{dt} \vec{p}_i\dmb
& = & - m\dmb \vec{u}_i^{(0)} ,
\\
\tau \frac{d}{dt} \vec{p}_i \rel
& = & - \Gamma m\dmb \Delta \vec{u}_i .
\Endeqa
We now expand $\vec{v}\sol$ by a Taylor series with respect to
$\vec{q}_i$. This can be formalized by introducing the operators
\Begeqa
{\Omega}_+ (\vec{q}) & = & 1 + 
\frac{1}{8} q_\alpha q_\beta \partial_\alpha \partial_\beta + \ldots ,
\\
{\Omega}_- (\vec{q}) & = & q_\alpha \partial_\alpha
+ \frac{1}{24}
q_\alpha q_\beta q_\gamma
\partial_\alpha \partial_\beta \partial_\gamma + \ldots ,
\Endeqa
which allows us to write
\Begeqa
\label{eq:uzeroParticle}
\vec{u}_i^{(0)}
& = &
\dot{\vec{r}}_i\dmb -
{\Omega}_+ (\vec{q_i}) \left. \vec{v}\sol (\vec{r})
\right\vert_{\vec{r} = \vec{r}_i\dmb} ,
\\
\label{eq:deltauParticle}
\Delta \vec{u}_i
& = &
\dot{\vec{q}}_i - 
{\Omega}_- (\vec{q_i}) \left. \vec{v}\sol (\vec{r})
\right\vert_{\vec{r} = \vec{r}_i\dmb} .
\Endeqa
We can now calculate the rate at which the microscopic
dumbbell Hamiltonian changes, again taking into account only
the dissipative part of the dynamics:
\Begeq
\frac{d}{dt} \hat{H}\dmb = \sum_i \left[
  \dot{\vec{r}}_i\dmb \cdot \dot{\vec{p}}_i\dmb +
  \dot{\vec{q}}_i \cdot \dot{\vec{p}}_i\rel \right] ,
\Endeq
\ie
\Begeq
\label{eq:DissipationRate1}
\tau \frac{d}{dt} \hat{H}\dmb = - m\dmb \sum_i \left[
  \dot{\vec{r}}_i\dmb \cdot \vec{u}_i^{(0)} +
  \Gamma \dot{\vec{q}}_i \cdot \Delta \vec{u}_i \right] .
\Endeq

Let us now consider the solvent. Again we ignore the Hamiltonian part
of the dynamics. We also ignore the contribution by viscous
dissipation ($\propto \eta_\abgd$), because this can be considered
separately from the dissipative coupling to the dumbbells --- it is
known that the viscous term is dissipative, conserves the momentum,
and just yields a well-known additive contribution to the overall
dissipation rate. We thus obtain
\Begeqa
\partial_t \rho\sol & = & 0 , \\
\partial_t \vec{j}\sol & = & \vec{f} ,
\Endeqa
with
\Begeqa
\vec f (\vec{r}) & = &
\sum_i \left[
  \delta (\vec{r} - \vec{r}_i\dmb - \vec{q}_i / 2) \vec{F}_i^{(1)}
  \right.
  \\
  \nonumber
  &&
  \left.
  +
  \delta (\vec{r} - \vec{r}_i\dmb + \vec{q}_i / 2) \vec{F}_i^{(2)}
  \right] ,
\\
\vec{F}_i^{(1)} & = & \zeta \vec{u}_i^{(1)} ,
\\
\vec{F}_i^{(2)} & = & \zeta \vec{u}_i^{(2)} .
\Endeqa
The rate of change of the solvent Hamiltonian, coming from the
dissipative coupling, is then calculated to be
\Begeqa
\frac{d}{dt} H\sol
& = &
\sum_i \left[
  \vec{v}\sol (\vec{r}_i\dmb + \vec{q}_i / 2)
  \cdot \vec{F}_i^{(1)}
  \right.
  \\
  \nonumber
  &&
  \left.
  +
  \vec{v}\sol (\vec{r}_i\dmb - \vec{q}_i / 2)
  \cdot \vec{F}_i^{(2)} \right] ,
\Endeqa
or
\Begeqa
\tau \frac{d}{dt} H\sol
& = &
m \sum_i \left[
  \vec{v}\sol (\vec{r}_i\dmb + \vec{q}_i / 2)
  \cdot \vec{u}_i^{(1)}
  \right.
  \\
  \nonumber
  &&
  \left.
  +
  \vec{v}\sol (\vec{r}_i\dmb - \vec{q}_i / 2)
  \cdot \vec{u}_i^{(2)} \right] .
\Endeqa
In terms of $\vec{u}_i^{(0)}$ and $\Delta \vec{u}_i$, this
is rewritten as
\Begeqa
\tau \frac{d}{dt} H\sol
& = &
m \sum_i \left[
  2 \vec{u}_i^{(0)} \cdot 
  {\Omega}_+ (\vec{q_i}) \left. \vec{v}\sol (\vec{r})
  \right\vert_{\vec{r} = \vec{r}_i\dmb}
  \right.
  \\
  \nonumber
  &&
  \left.
  +
  \frac{1}{2} \Delta \vec{u}_i \cdot
  {\Omega}_- (\vec{q_i}) \left. \vec{v}\sol (\vec{r})
  \right\vert_{\vec{r} = \vec{r}_i\dmb} \right] ,
\Endeqa
or, taking into account $2 m = m\dmb$, $\Gamma = 1/4$,
\Begeqa
\label{eq:DissipationRate2}
\tau \frac{d}{dt} H\sol
& = &
m\dmb \sum_i \left[
  \vec{u}_i^{(0)} \cdot 
  {\Omega}_+ (\vec{q_i}) \left. \vec{v}\sol (\vec{r})
  \right\vert_{\vec{r} = \vec{r}_i\dmb}
  \right.
  \\
  \nonumber
  &&
  \left.
  +
  \Gamma \Delta \vec{u}_i \cdot
  {\Omega}_- (\vec{q_i}) \left. \vec{v}\sol (\vec{r})
  \right\vert_{\vec{r} = \vec{r}_i\dmb} \right] .
\Endeqa
Combining Eqs.~\ref{eq:DissipationRate1} and \ref{eq:DissipationRate2},
we find for the total dissipation rate that comes from the dumbbell-solvent
coupling in the microscopic model:
\Begeqa
&&
\tau \frac{d}{dt} \left( \hat{H}\dmb + H\sol \right)
\\
\nonumber
& = &
- m\dmb \sum_i \left[ \left( \vec{u}_i^{(0)} \right)^2 +
  \Gamma \left( \Delta \vec{u}_i \right)^2 \right] ,
\Endeqa
showing that the coupling is strictly compatible with the Second Law.

On the field-theoretic level, we postulate the analogous expression
\Begeqa
&&
\tau \frac{d}{dt} \left( H\dmb + H\sol \right)
\\
\nonumber
& = &
- \int d^3 \vec{r} \rho\dmb (\vec{r}) \left[
  \left( \vec{u}^{(0)} (\vec{r}) \right)^2 +
  \Gamma \left( \Delta \vec{u} (\vec{r}) \right)^2 \right]
\Endeqa
with the field analogues of Eqs.~\ref{eq:uzeroParticle} and
\ref{eq:deltauParticle},
\Begeqa
\vec{u}^{(0)} (\vec{r})
& = &
\vec{v}\dmb (\vec{r}) -
{\Omega}_+ (\vec{q} (\vec{r})) \vec{v}\sol (\vec{r}) ,
\\
\Delta \vec{u} (\vec{r})
& = &
\vec{v}\rel (\vec{r}) -
{\Omega}_- (\vec{q} (\vec{r})) \vec{v}\sol (\vec{r}) .
\Endeqa
Decomposing the dissipation rate into the various contributions,
we find
\Begeqa
&& - \tau \frac{d}{dt} \left( H\dmb + H\sol \right)
\\
\nonumber
& = &
\int d^3 \vec{r} \rho\dmb
\left( {\Omega}_+ \vec{v}\sol \right)^2
+
\Gamma \int d^3 \vec{r} \rho\dmb
\left( {\Omega}_- \vec{v}\sol \right)^2
\\
\nonumber
& & + 
\int d^3 \vec{r} \rho\dmb \left( \vec{v}\dmb \right)^2
+
\Gamma \int d^3 \vec{r} \rho\dmb \left( \vec{v}\rel \right)^2
\\
\nonumber
& & -
2 \int d^3 \vec{r} \rho\dmb
\vec{v}\dmb \cdot {\Omega}_+ \vec{v}\sol
\\
\nonumber
& & -
2 \Gamma \int d^3 \vec{r} \rho\dmb
\vec{v}\rel \cdot {\Omega}_- \vec{v}\sol .
\Endeqa
At this point, we introduce the adjoint operators
\Begeqa
{\Omega}_+^\dagger & = & 1 + 
\frac{1}{8} \partial_\alpha \partial_\beta q_\alpha q_\beta + \ldots ,
\\
{\Omega}_-^\dagger & = & - \partial_\alpha q_\alpha
- \frac{1}{24}
\partial_\alpha \partial_\beta \partial_\gamma
q_\alpha q_\beta q_\gamma - \ldots ;
\Endeqa
this allows us to write
\Begeqa
&& - \tau \frac{d}{dt} \left( H\dmb + H\sol \right)
\\
\nonumber
& = &
\int d^3 \vec{r} \vec{v}\sol \cdot \left[
  {\Omega}_+^\dagger \rho\dmb {\Omega}_+
  + \Gamma
  {\Omega}_-^\dagger \rho\dmb {\Omega}_-
  \right] \vec{v}\sol
\\
\nonumber
& - &
\int d^3 \vec{r} \vec{v}\sol \cdot
{\Omega}_+^\dagger \rho\dmb \vec{v}\dmb
- \Gamma
\int d^3 \vec{r} \vec{v}\sol \cdot
{\Omega}_-^\dagger \rho\dmb \vec{v}\rel
\\
\nonumber
& + &
\int d^3 \vec{r} \vec{v}\dmb \cdot \rho\dmb \vec{v}\dmb
-
\int d^3 \vec{r} \vec{v}\dmb \cdot
\rho\dmb {\Omega}_+ \vec{v}\sol
\\
\nonumber
& + &
\Gamma \int d^3 \vec{r} \vec{v}\rel \cdot \rho\dmb \vec{v}\rel 
- \Gamma \int d^3 \vec{r} \vec{v}\rel \cdot
\rho\dmb {\Omega}_- \vec{v}\sol  .
\Endeqa
Recalling $\vec{v}\sol = \delta H / \delta \vec{j}\sol$, $\vec{v}\dmb
= \delta H / \delta \vec{j}\dmb$, and $\vec{v}\rel = \delta H / \delta
\vec{j}\rel$, one sees (cf. Eq.~\ref{eq:GENERICDissipationRate}) that
this is precisely the form that is required by the general
formalism. This, in turn, allows us to directly read off the continuum
equations of motion (again, we emphasize that we here ignore the
Hamiltonian contribution and the viscous part $\propto \eta_\abgd$):
\Begeqa
\partial_t \rho\sol & = & 0 , \\
\partial_t \rho\dmb & = & 0 , \\
\partial_t \vec{k}\rel & = & 0 , \\
\tau \partial_t \vec{j}\sol & = &
{\Omega}_+^\dagger \rho\dmb
\left[ \vec{v}\dmb - {\Omega}_+ \vec{v}\sol \right]
\\
\nonumber
&&
+ \Gamma {\Omega}_-^\dagger \rho\dmb
\left[ \vec{v}\rel - {\Omega}_- \vec{v}\sol \right] , \\
\tau \partial_t \vec{j}\dmb & = &
- \rho\dmb \left[ \vec{v}\dmb - {\Omega}_+ \vec{v}\sol \right] , \\
\tau \partial_t \vec{j}\rel & = &
- \Gamma \rho\dmb \left[
  \vec{v}\rel - {\Omega}_- \vec{v}\sol \right] .
\Endeqa

One can show that these equations are compatible with momentum
conservation. To this end, we note that Gauss' theorem implies
\Begeqa
\int d^3\vec{r} {\Omega}_+^\dagger (\ldots) & = &
\int d^3\vec{r} 1 (\ldots) , \\
\int d^3\vec{r} {\Omega}_-^\dagger (\ldots) & = & 0 .
\Endeqa
Therefore
\Begeqa
&&
\tau \frac{d}{dt} \int d^3 \vec{r} \vec{j}\sol
\\
\nonumber
& = &
\int d^3 \vec{r} \rho\dmb \vec{v}\dmb 
- \int d^3 \vec{r} \rho\dmb {\Omega}_+ \vec{v}\sol ,
\\
&&
\tau \frac{d}{dt} \int d^3 \vec{r} \vec{j}\dmb
\\
\nonumber
& = &
- \int d^3 \vec{r} \rho\dmb \vec{v}\dmb
+ \int d^3 \vec{r} \rho\dmb {\Omega}_+ \vec{v}\sol ,
\Endeqa
\Begeq
\tau \frac{d}{dt} \int d^3 \vec{r}
\left( \vec{j}\sol + \vec{j}\dmb \right) = 0 ;
\Endeq
the last equation obviously implies conservation of the total momentum
--- note that the relative motion of the beads with respect to each
other does not contribute to the overall momentum balance.

\section{Final set of equations}
\label{sec:FinalSet}

Let us summarize what we have achieved so far. We started from an
extremely simple microscopic dumbbell model for viscoelastic phase
separation. We then defined a set of fields (two scalar fields, four
vector fields) which can be explicitly constructed from the
microscopic configurations in real and momentum space. The microscopic
dynamics then allowed us to evaluate the Poisson brackets of the
fields; from there it turned out that the field-theoretic set of
variables is closed, meaning that the Poisson brackets do not generate
additional variables. We then applied four important approximations
and assumptions, which allowed us to go from the microscopic model to
field theory. These are: (i) Replacement of the non-bonded
interactions with a Van der Waals free energy, augmented with an
interfacial stiffness term; (ii) replacement of the particle
Hamiltonian with the corresponding field-theoretic expression; (iii) a
similar replacement for the dissipation rate, and (iv) the assumption
that the molecular conformations may be represented by a smooth vector
field. \emph{No further assumptions or approximations were made}. The
formalism of dissipative Hamiltonian systems then made it possible to
construct the field-theoretic equations of motion in a somewhat
tedious but straightforward fashion. We thus arrive at a set of
equations which are fully compatible with non-equilibrium
thermodynamics, conserve the momentum, and have a well-defined
transformation behavior under flip (which is easily checked by
inspection). In total, the resulting equations read:
\Begeqa
\partial_t \rho\sol + \nabla \cdot \vec{j}\sol & = & 0 ,
\\
\partial_t \rho\dmb + \nabla \cdot \vec{j}\dmb & = & 0 ,
\Endeqa
\Begeq
\partial_t \vec{k}\rel +
\nabla \cdot \left( \vec{k}\rel \vec{v}\dmb \right)
=
\frac{k}{m\rel} \vec{j}\rel ,
\Endeq
\Begeqa
&&
\partial_t \vec{j}\sol +
\nabla \cdot \left( \vec{j}\sol \vec{v}\sol \right)
\\
\nonumber
& = &
- \nabla p\sol
+ \boldsymbol{\eta} \cdddot \nabla \nabla \vec{v}\sol
+ \frac{1}{\tau}
{\Omega}_+^\dagger \rho\dmb
\left[ \vec{v}\dmb - {\Omega}_+ \vec{v}\sol \right]
\\
\nonumber
&&
+ \frac{\Gamma}{\tau} {\Omega}_-^\dagger \rho\dmb
\left[ \vec{v}\rel - {\Omega}_- \vec{v}\sol \right] ,
\\
&&
\partial_t \vec{j}\dmb +
\nabla \cdot \left( \vec{j}\dmb \vec{v}\dmb \right)
\\
\nonumber
& = &
- \nabla p\dmb +
\kappa \rho\dmb \nabla \nabla^2 \rho\dmb
% \\
% \nonumber
% &&
- \frac{1}{\tau} \rho\dmb \left[
  \vec{v}\dmb - {\Omega}_+ \vec{v}\sol \right] ,
\\
&&
\partial_t \vec{j}\rel + \nabla \cdot \left(
\vec{j}\rel \vec{v}\dmb \right)
\\
\nonumber
& = &
- \vec{k}\rel
- \frac{\Gamma}{\tau} \rho\dmb \left[
  \vec{v}\rel - {\Omega}_- \vec{v}\sol \right] .
\Endeqa

By making use of the mass conservation equations, we can transform
these equations for ``extensive'' fields ($\vec{k}\rel$,
$\vec{j}\sol$, $\vec{j}\dmb$, $\vec{j}\rel$) to equivalent equations
for the corresponding ``intensive'' fields ($\vec{q}$, $\vec{v}\sol$,
$\vec{v}\dmb$, $\vec{v}\rel$). Here it is useful to introduce the
convective derivatives
\Begeqa
D_t\sol & = & \partial_t + \vec{v}\sol \cdot \nabla , \\
D_t\dmb & = & \partial_t + \vec{v}\dmb \cdot \nabla .
\Endeqa
Straightforward transformation yields
\Begeqa
\label{eq:FullEqMotion1}
D_t\sol \rho\sol + \rho\sol \nabla \cdot \vec{v}\sol & = & 0 ,
\\
\label{eq:FullEqMotion2}
D_t\dmb \rho\dmb + \rho\dmb \nabla \cdot \vec{v}\dmb & = & 0 ,
\Endeqa
\Begeq
\label{eq:FullEqMotion3}
D_t\dmb \vec{q} = \vec{v}\rel ,
\Endeq
\Begeqa
\label{eq:FullEqMotion4}
\rho\sol D_t\sol \vec{v}\sol
& = &
- \nabla p\sol
+ \boldsymbol{\eta} \cdddot \nabla \nabla \vec{v}\sol
\\
\nonumber
&&
+ \frac{1}{\tau}
{\Omega}_+^\dagger \rho\dmb
\left[ \vec{v}\dmb - {\Omega}_+ \vec{v}\sol \right]
\\
\nonumber
&&
+ \frac{\Gamma}{\tau} {\Omega}_-^\dagger \rho\dmb
\left[ \vec{v}\rel - {\Omega}_- \vec{v}\sol \right] ,
\\
\label{eq:FullEqMotion5}
\rho\dmb D_t\dmb \vec{v}\dmb
& = &
- \nabla p\dmb +
\kappa \rho\dmb \nabla \nabla^2 \rho\dmb
\\
\nonumber
&&
- \frac{1}{\tau} \rho\dmb \left[
  \vec{v}\dmb - {\Omega}_+ \vec{v}\sol \right] ,
\Endeqa
\Begeq
\label{eq:FullEqMotion6}
D_t\dmb \vec{v}\rel = - \frac{k}{m\rel} \vec{q}
- \frac{1}{\tau} \left[
  \vec{v}\rel - {\Omega}_- \vec{v}\sol \right] .
\Endeq

It should be noted that for the case $\vec{q} = 0$, $\vec{v}\rel = 0$
(which implies ${\Omega}_+ = 1$, ${\Omega}_- = 0$)
we recover a simple non-viscoelastic two-fluid model:
\Begeqa
D_t\sol \rho\sol + \rho\sol \nabla \cdot \vec{v}\sol & = & 0 ,
\\
D_t\dmb \rho\dmb + \rho\dmb \nabla \cdot \vec{v}\dmb & = & 0 ,
\Endeqa
\Begeqa
\rho\sol D_t\sol \vec{v}\sol
& = &
- \nabla p\sol
+ \boldsymbol{\eta} \cdddot \nabla \nabla \vec{v}\sol
\\
\nonumber
&&
+ \frac{1}{\tau} \rho\dmb
\left[ \vec{v}\dmb - \vec{v}\sol \right] ,
\\
\rho\dmb D_t\dmb \vec{v}\dmb
& = &
- \nabla p\dmb +
\kappa \rho\dmb \nabla \nabla^2 \rho\dmb
\\
\nonumber
&&
- \frac{1}{\tau} \rho\dmb \left[
  \vec{v}\dmb - \vec{v}\sol \right] .
\Endeqa

\section{Transformation to new variables}
\label{sec:transformation}

It is instructive to transform the equations to a new set of variables
that is adapted to the kinematics of the two-body problem. We start by
defining the total mass density
\Begeq
\rho = \rho\dmb + \rho\sol
\Endeq
and the reduced mass density
\Begeq
\rho\red = \rho^{-1} \rho\dmb \rho\sol .
\Endeq
In velocity space, we introduce the mass-averaged velocity
\Begeq
\vec{V} = \rho^{-1} \left( \rho\dmb \vec{v}\dmb
+ \rho\sol \vec{v}\sol \right)
\Endeq
and the relative velocity
\Begeq
\vec{w} = \vec{v}\dmb - \vec{v}\sol .
\Endeq
The inverse transformation is given by
\Begeqa
\vec{v}\dmb & = & \vec{V} + \frac{\rho\red}{\rho\dmb} \vec{w} , \\
\vec{v}\sol & = & \vec{V} - \frac{\rho\red}{\rho\sol} \vec{w} .
\Endeqa
Furthermore, we define a new convective derivative via
\Begeq
D_t = \partial_t + \vec{V} \cdot \nabla ,
\Endeq
such that
\Begeqa
D_t\dmb & = & D_t + \frac{\rho\red}{\rho\dmb} \vec{w} \cdot \nabla , \\
D_t\sol & = & D_t - \frac{\rho\red}{\rho\sol} \vec{w} \cdot \nabla .
\Endeqa
The mass conservation equations for $\rho\dmb$ and $\rho\sol$
are thus written as
\Begeqa
\label{eq:EqMotionRhoDumb}
D_t \rho\dmb + \rho\dmb \nabla \cdot \vec{V}
+ \nabla \cdot \left( \rho\red \vec{w} \right) & = & 0 , \\
D_t \rho\sol + \rho\sol \nabla \cdot \vec{V}
- \nabla \cdot \left( \rho\red \vec{w} \right) & = & 0 .
\Endeqa
For the total mass density this implies the simple relation
\Begeq
D_t \rho + \rho \nabla \cdot \vec{V} = 0 .
\Endeq
Apart from $\rho$, we need yet another combination of $\rho\dmb$ and
$\rho\sol$ to describe the dynamics of the density
contrast. $\rho\red$ is not suitable for that purpose, due to its
invariance with respect to the exchange $\rho\dmb \leftrightarrow
\rho\sol$. We therefore take the normalized density difference
\Begeq
c = \rho^{-1} \left( \rho\dmb - \rho\sol \right) .
\Endeq
This implies
\Begeqa
\rho\dmb & = & \frac{\rho}{2} (1 + c) , \\
\rho\sol & = & \frac{\rho}{2} (1 - c) , \\
\rho\red & = & \frac{\rho}{4} \left( 1 - c^2 \right) , \\
\vec{v}\dmb & = & \vec{V} + \frac{1}{2} (1 - c) \vec{w} , \\
\vec{v}\sol & = & \vec{V} - \frac{1}{2} (1 + c) \vec{w} , \\
D_t\dmb & = & D_t + \frac{1}{2} (1 - c) \vec {w} \cdot \nabla , \\
D_t\sol & = & D_t - \frac{1}{2} (1 + c) \vec {w} \cdot \nabla .
\Endeqa

From the equations of motion for $\rho\dmb$ and $\rho\sol$ we can find
the equations of motion for $c$ and $\rho\red$:
\Begeq
\rho D_t c = - 2 \, \nabla \cdot \left( \rho\red \vec{w} \right) ,
\Endeq
\Begeq
\partial_t \rho\red + \nabla \cdot \left( \rho\red \vec{V} \right)
= c \, \nabla \cdot \left( \rho\red \vec{w} \right) .
\Endeq

We now turn to the velocity equations. We already have derived
the dynamics for $\vec{v}\dmb$ and $\vec{v}\sol$, which we
abbreviate as
\Begeqa
\rho\dmb D_t\dmb \vec{v}\dmb & = & \vec{f}\dmb , \\
\rho\sol D_t\sol \vec{v}\sol & = & \vec{f}\sol .
\Endeqa
This information, together with the equations of motion for the
densities, is sufficient to construct the equations of motion for
$\vec{V}$ and $\vec{w}$. After some lengthy algebra we finally find
\Begeq
\label{eq:EqMotionV}
\rho D_t \vec{V} + \nabla \cdot
\left( \rho\red \vec{w} \vec{w} \right)
=
\vec{f}\dmb + \vec{f}\sol ,
\Endeq
\Begeqa
\label{eq:EqMotionw}
&&
D_t \vec{w} + \vec{w} \cdot \nabla \vec{V}
- c \vec{w} \cdot \nabla \vec{w}
- \frac{1}{2} \vec{w} \vec{w} \cdot \nabla c
\\
\nonumber
& = &
\frac{1}{\rho\dmb} \vec{f}\dmb -
\frac{1}{\rho\sol} \vec{f}\sol .
\Endeqa

The equations for $\vec{q}$ and $\vec{v}\rel$ may also be transformed;
however, we believe this does not provide lots of insight. For this
reason, we do not mention the explicit expressions here.

\section{Approximations}
\label{sec:approx}

We now subject the derived equations of motion to a number of
approximations, and by this try to find guidelines to construct a
simplified field-theoretic model.
\begin{enumerate}

\item \emph{Overdamped harmonic oscillator.} We assume that
  inertial effects for the motion of $\vec{q}$ are negligible.
  This is a standard assumption in the theory of polymer dynamics,
  e.g. the Rouse model (see, e.g., Ref.~\cite{doi_theory_1988}).
  This means that in Eq.~\ref{eq:FullEqMotion6} we set
  $D_t\dmb \vec{v}\rel = 0$. This yields
  \Begeq
  \frac{1}{\tau} \left[
    \vec{v}\rel - {\Omega}_- \vec{v}\sol \right]
  = - \frac{k}{m\rel} \vec{q}
  \Endeq
  and
  \Begeq
  \vec{v}\rel = {\Omega}_- \vec{v}\sol
  - \frac{k \tau}{m\rel} \vec{q}
  = {\Omega}_- \vec{v}\sol - \frac{1}{\tau_q} \vec{q} ;
  \Endeq 
  here we have introduced the relaxation time $\tau_q = m\rel / (k
  \tau)$, which may be viewed as the configurational relaxation time
  of the polymer chains. We may thus write $k / m\rel = 1 / (\tau \tau_q)$;
  note that we should view $\tau$ as a rather small time and $\tau_q$
  as a large time such that the product $\tau \tau_q$ is of order unity.

  In the other equations, we thus eliminate $\vec{v}\rel$. On the one
  hand, we obtain a first-order equation of motion for $\vec{q}$:
  \Begeq
  D_t\dmb \vec{q} = {\Omega}_- \vec{v}\sol
  - \frac{1}{\tau_q}  \vec{q} ,
  \Endeq
  while on the other hand the force expression for the solvent
  is simplified:
  \Begeqa
  \vec{f}\sol
  & = &
  - \nabla p\sol
  + \boldsymbol{\eta} \cdddot \nabla \nabla \vec{v}\sol
  \\
  \nonumber
  & + & \frac{1}{\tau}
  {\Omega}_+^\dagger \rho\dmb
  \left[ \vec{v}\dmb - {\Omega}_+ \vec{v}\sol \right]
  - \frac{\Gamma}{\tau \tau_q}
  {\Omega}_-^\dagger \rho\dmb \vec{q} .
  \Endeqa

\item \emph{Lowest-order viscoelastic coupling.} The operators
  ${\Omega}_+$ and ${\Omega}_-$ represent the
  Taylor expansion of the flow field on the scale of the extension
  of the macromolecules. It is reasonable to assume that the flow
  field does not vary extremely strongly on that scale, such that
  a low-order Taylor expansion should be sufficient. We here assume
  that actually an expansion up to linear order is good enough,
  which means that we set ${\Omega}_- = \vec{q} \cdot \nabla$
  and ${\Omega}_+ = 1$. This simplifies the force expressions
  significantly:
  \Begeqa
  \vec{f}\sol
  & = &
  - \nabla p\sol
  + \boldsymbol{\eta} \cdddot \nabla \nabla \vec{v}\sol
  \\
  \nonumber
  && 
  + \frac{1}{\tau} \rho\dmb \vec{w}
  + \frac{\Gamma}{\tau \tau_q}
  \nabla \cdot \left( \rho\dmb \vec{q} \vec{q} \right) , 
  \\
  \vec{f}\dmb
  & = &
  - \nabla p\dmb +
  \kappa \rho\dmb \nabla \nabla^2 \rho\dmb
  - \frac{1}{\tau} \rho\dmb \vec{w} ,
  \Endeqa
  which means ($p = p\sol + p\dmb$ denotes the total pressure)
  \Begeqa
  &&
  \vec{f}\dmb + \vec{f}\sol =
  - \nabla p
  + \boldsymbol{\eta} \cdddot \nabla \nabla \vec{v}\sol
  \\
  \nonumber
  &&
  + \kappa \rho\dmb \nabla \nabla^2 \rho\dmb
  + \frac{\Gamma}{\tau \tau_q}
  \nabla \cdot \left( \rho\dmb \vec{q} \vec{q} \right) ,
  \Endeqa
  \Begeqa
  \label{eq:DeltaForceTerm}
  &&
  \frac{1}{\rho\dmb} \vec{f}\dmb - \frac{1}{\rho\sol} \vec{f}\sol
  =
  - \frac{\nabla p\dmb}{\rho\dmb} + \frac{\nabla p\sol}{\rho\sol}
  \\
  \nonumber
  &&
  - \frac{\boldsymbol{\eta}}{\rho\sol} \cdddot \nabla \nabla \vec{v}\sol
  + \kappa \nabla \nabla^2 \rho\dmb
  - \frac{1}{\tau} \frac{2}{1 - c} \vec{w}
  \\
  \nonumber
  &&
  - \frac{\Gamma}{\tau \tau_q} \frac{1}{\rho\sol} 
  \nabla \cdot \left( \rho\dmb \vec{q} \vec{q} \right) .
  \Endeqa
  The equation of motion for $\vec{q}$ gets simplified further:
  \Begeq
  \label{eq:FirstEqMotionQ}
  D_t\dmb \vec{q} = \vec{q} \cdot \nabla \vec{v}\sol
  - \frac{1}{\tau_q} \vec{q} .
  \Endeq

\item \emph{Incompressibility.} We assume that the total mass density
  is spatially and temporally constant. The conservation equation for
  the total mass then simplifies to
  \Begeq
  \nabla \cdot \vec{V} = 0.
  \Endeq
  The pressure is therefore no longer derived from an equation of
  state, but rather acts as a Lagrange multiplier to enforce the
  incompressibility constraint.

\item \emph{Small $\vec{w}$.} The relative velocity $\vec{w}$ is a
  non-hy\-drodynamic variable, which is therefore expected to relax
  fairly rapidly. Therefore it is expected to never deviate very much
  from its value at local thermal equilibrium, which is zero. Let us
  therefore inspect the so-far derived dynamics for terms linear or
  quadratic in $\vec{w}$. For this purpose, we replace $\vec{w} \to
  \varepsilon \vec{w}$, where $\varepsilon$ is a scalar expansion
  parameter. This allows us to sort the expressions in terms of powers
  of $\varepsilon$.

  Firstly, we have the conservation of total mass,
  \Begeq
  \nabla \cdot \vec{V} = 0 ,
  \Endeq
  and, secondly, the conservation of the composition, which we
  represent by the dynamics for $\rho\dmb$:
  \Begeq
  D_t \rho\dmb = O(\varepsilon) .
  \Endeq
  Thirdly, we have the overdamped dynamics for $\vec{q}$:
  \Begeq
  D_t \vec{q} = \vec{q} \cdot \nabla \vec{V}
  - \frac{1}{\tau_q} \vec{q} + O(\varepsilon) .
  \Endeq
  In the fourth place, we need to consider the momentum balance
  \Begeqa
  &&
  \rho D_t \vec{V}
  \\
  \nonumber
  & = &
  - \nabla p
  + \boldsymbol{\eta} \cdddot \nabla \nabla \vec{V}
  + \kappa \rho\dmb \nabla \nabla^2 \rho\dmb
  \\
  \nonumber
  &&
  + \frac{\Gamma}{\tau \tau_q}
  \nabla \cdot \left( \rho\dmb \vec{q} \vec{q} \right)
  + O(\varepsilon)
  \\
  \nonumber
  & = &
  - \nabla p
  + \eta_{\mathrm{s}} \nabla^2 \vec{V}
  + \kappa \rho\dmb \nabla \nabla^2 \rho\dmb
  \\
  \nonumber
  &&
  + \frac{\Gamma}{\tau \tau_q}
  \nabla \cdot \left( \rho\dmb \vec{q} \vec{q} \right)
  + O(\varepsilon) .
  \Endeqa

  And finally we need to consider the equation of motion for
  $\vec{w}$ (cf. Eq.~\ref{eq:EqMotionw}), which we do not
  write down explicitly here.

\item \emph{Overdamped dynamics for $\vec{w}$.} It is assumed that
  $\vec{w}$ is a fast variable and that it may therefore be
  adiabatically eliminated for time scales significantly larger than
  $\tau$, similar to the adiabatic elimination of $\vec{v}\rel$ at the
  beginning of this section. To do this systematically is however a
  daunting task, and probably (if possible and successful) only of
  limited value, since the resulting set is probably not fully
  consistent with non-equilibrium thermodynamics. We therefore take a
  simpler approach and rather replace the terms $O(\varepsilon)$ with
  unknown phenomenological terms which need to be chosen in order to
  ensure consistency with non-equilibrium thermodynamics.
\end{enumerate}

Based upon this philosophy, we thus obtain a set of dynamic equations
for the three fields $\rho\dmb$, $\vec{V}$, and $\vec{q}$. They read:
\Begeqa
\label{eq:FinalApprox1}
\nabla \cdot \vec{V} & = & 0 ,
\\
\label{eq:FinalApprox2}
D_t \rho\dmb & = & - \nabla \cdot \vec{j}\nt ,
\\
\label{eq:FinalApprox3}
\rho D_t \vec{V} & = & - \nabla p
+ \eta_{\mathrm{s}} \nabla^2 \vec{V}
+ \kappa \rho\dmb \nabla \nabla^2 \rho\dmb
\\
\nonumber
& & 
+ \frac{\Gamma}{\tau \tau_q}
\nabla \cdot \left( \rho\dmb \vec{q} \vec{q} \right)
+ \nabla \cdot \boldsymbol{\sigma} ,
\\
\label{eq:FinalApprox4}
D_t \vec{q} & = & \vec{q} \cdot \nabla \vec{V}
  - \frac{1}{\tau_q} \vec{q} + \vec{Q} .
\Endeqa
Here we have introduced three phenomenological terms: (i) the
interdiffusion current $\vec{j}\nt$, (ii) a stress tensor
$\boldsymbol{\sigma}$, and (iii) the vector $\vec{Q}$, which describes
the influence of $\vec{w}$ on the dynamics of $\vec{q}$. Note that
divergence operators have been introduced in order to keep the
conservation laws for the dumbbell mass and the overall momentum.

We have thus gone from a set for two scalar fields and four vector
fields to a simplified set that involves only one scalar field and two
vector fields as the system's state variables, plus the pressure that
acts as a Lagrange multiplier for incompressibility. The arguments
presented in this section should not be viewed as rigorous but rather
as heuristic. It should be noted that the term $\nabla \cdot \left(
\rho\dmb \vec{q} \vec{q} \right)$ is symmetric under time reversal and
hence conservative. The prefactor has been written in terms of two
relaxation times; however, physically this should be viewed as the
square of an oscillation frequency.

\section{Simplified model}
\label{sec:simplified}

It is natural to ask to what extent Poisson brackets might be helpful
in deriving Eqs.~\ref{eq:FinalApprox1}-\ref{eq:FinalApprox4}. To
answer this question, let us consider a compressible system (we do not
wish to deal with the mathematical complications that arise from an
incompressibility constraint), where the dynamical variables are the
fields $\rho$, $\rho\dmb$, $\vec{j} = \rho \vec{V}$ and $\vec{q}$ and
Hamiltonian
\Begeq
H = \int d^3\vec{r} \left[ \frac{\vec{j}^2}{2 \rho} + f
  + \frac{\kappa}{2} \left( \nabla \rho\dmb \right)^2
  + \frac{k}{2} \frac{\rho\dmb}{m\dmb} \vec{q}^2 \right] ;
\Endeq
here $f$ is the free energy density depending on both $\rho$
and $\rho\dmb$.

For the non-vanishing Poisson brackets we find
\Begeq
\mybrack{ \rho (\vec{r}), \vec{j} (\vec{r}') }
=  - \rho (\vec{r}') \nabla \delta (\vec{r} - \vec{r}') ,
\Endeq
\Begeq
\mybrack{ \rho\dmb (\vec{r}), \vec{j} (\vec{r}') }
= - \rho\dmb (\vec{r}') \nabla \delta (\vec{r} - \vec{r}') ,
\Endeq
\Begeqa
&&
\mybrack{q_\alpha (\vec{r}), j_\beta (\vec{r}') }
\\
\nonumber
& = &
\left[ \frac{\rho\dmb (\vec{r}')}{\rho\dmb (\vec{r})} q_\alpha (\vec{r})
  - q_\alpha (\vec{r}') \right] \partial_\beta \delta (\vec{r} - \vec{r}') ,
\Endeqa
\Begeqa
&&
\mybrack{ j_\alpha (\vec{r}), j_\beta (\vec{r}') }
\\
\nonumber
& = &
j_\beta (\vec{r}) \partial_\alpha' \delta ( \vec{r}' - \vec{r}) -
j_\alpha (\vec{r}') \partial_\beta \delta ( \vec{r} - \vec{r}') ,
\Endeqa
while the functional derivatives of the Hamiltonian are
\Begeqa
\frac{\delta H}{\delta \vec{j}} & = & \vec{V} ,
\\
\frac{\delta H}{\delta \rho} & = & - \frac{1}{2} \vec{V}^2 +
\frac{\partial f}{\partial \rho} ,
\\
\frac{\delta H}{\delta \rho\dmb} & = &
\frac{\partial f}{\partial \rho\dmb} - \kappa \nabla^2 \rho\dmb
+ \frac{k}{2 m\dmb} \vec{q}^2 ,
\\
\frac{\delta H}{\delta \vec{q}} & = &
k \frac{\rho\dmb}{m\dmb} \vec{q} .
\Endeqa
We can now insert these results to calculate the conservative equations
of motion, \emph{as they are produced by the Poisson bracket
  formalism}. After some lengthy but straightforward algebra we find
\Begeqa
\partial_t \rho & = & - \nabla \cdot \left( \rho \vec V \right) ,
\\
\partial_t \rho\dmb & = & - \nabla \cdot \left( \rho\dmb \vec V \right) ,
\\
D_t q_\alpha & = & q_\alpha \frac{1}{\rho\dmb} \vec{V} \cdot \nabla \rho\dmb ,
\\
\partial_t j_\alpha & = & - \partial_\beta \left( j_\alpha V_\beta \right)
- \partial_\alpha p + \kappa \rho\dmb \nabla^2 \rho\dmb
\\
\nonumber
&&
- \frac{k}{m\dmb} \vec{q}^2 \partial_\alpha \rho\dmb ;
\Endeqa
here the pressure involves contributions from both $\partial f /
\partial \rho$ and $\partial f / \partial \rho\dmb$. Comparing this
with the set of equations that we heuristically derived in the
previous section, we see that (i) the equation of motion for $\vec{q}$
couples to the flow field in a significantly different fashion, and
that (ii) the elastic force term in the momentum equation looks
different, and \emph{does not conserve the momentum}. Such an equation
is however simply unsuitable for hydrodynamics. We therefore conclude
that for the simplified (or reduced) set of equations of motion
\emph{we have to abandon the Poisson bracket formalism}. In other
words: It is impossible to eliminate the undesired velocities
adiabatically, and at the same time maintain the Hamiltonian structure
of the theory. Given the fact that momentum variables play a decisive
role in Hamiltonian dynamics, and the fact that we have removed them,
this result is hardly surprising. It is perhaps possible to do the
development within some Hamiltonian formalism with constraints, but
this is beyond the scope of the present paper.

We therefore go back to
Eqs.~\ref{eq:FinalApprox1}-\ref{eq:FinalApprox4}, which we consider as
a reasonable starting point for further developments. The Hamiltonian
\Begeqa
\nonumber
H & = & \int d^3\vec{r} \left[ \frac{\rho}{2} \vec{V}^2 + f
  + \frac{\kappa}{2} \left( \nabla \rho\dmb \right)^2
  + \frac{k}{2} \frac{\rho\dmb}{m\dmb} \vec{q}^2 \right]
\\
\nonumber
& = &
\int d^3\vec{r} \left[ \frac{\rho}{2} \vec{V}^2 + f
  + \frac{\kappa}{2} \left( \nabla \rho\dmb \right)^2
  \right.
  \\
  &&
  \left.
  + \frac{1}{2} \frac{\Gamma}{\tau \tau_q} \rho\dmb \vec{q}^2 \right]
\Endeqa
must be viewed as a functional of $\vec{V}$, $\vec{q}$ and
$\rho\dmb$; its derivatives are
\Begeqa
\frac{\delta H}{\delta \vec{V}} & = & \rho \vec{V} ,
\\
\label{eq:FunctDerivRhoDumb}
\frac{\delta H}{\delta \rho\dmb} & = &
\frac{\partial f}{\partial \rho\dmb} - \kappa \nabla^2 \rho\dmb
+ \frac{1}{2} \frac{\Gamma}{\tau \tau_q} \vec{q}^2 ,
\\
\frac{\delta H}{\delta \vec{q}} & = &
\frac{\Gamma}{\tau \tau_q} \rho\dmb \vec{q} .
\Endeqa

We now consider a reduced version of the dynamic equations
Eq.~\ref{eq:FinalApprox1}-\ref{eq:FinalApprox4}, where we discard the
phenomenological terms $\vec{j}\nt$, $\boldsymbol{\sigma}$ and $\vec{Q}$,
as well as the viscous dissipation and the dissipative relaxation
of $\vec{q}$ ($\propto 1 / \tau_q$):
\Begeqa
\label{eq:FinalApprox1Hamil}
\nabla \cdot \vec{V} & = & 0 ,
\\
\label{eq:FinalApprox2Hamil}
D_t \rho\dmb & = & 0 ,
\\
\label{eq:FinalApprox3Hamil}
\rho D_t \vec{V} & = & - \nabla p
+ \kappa \rho\dmb \nabla \nabla^2 \rho\dmb
\\
\nonumber
&&
+ \frac{\Gamma}{\tau \tau_q}
\nabla \cdot \left( \rho\dmb \vec{q} \vec{q} \right)
\\
\label{eq:FinalApprox4Hamil}
D_t \vec{q} & = & \vec{q} \cdot \nabla \vec{V} .
\Endeqa

In what follows, we wish to demonstrate that this system is conservative,
\ie is characterized by $d H / dt = 0$.

We first notice that for any field $\phi$ the relation
\Begeq
\int d^3\vec{r} \phi \vec{V} \cdot \nabla \phi = 0
\Endeq
holds, as can be seen from integration by parts and incompressibility.
As a consequence, we have
\Begeq
\label{eq:UsefulDerivativeTransformation}
\int d^3\vec{r} \phi D_t \phi =
\int d^3\vec{r} \phi \partial_t \phi .
\Endeq
From Eq.~\ref{eq:FunctDerivRhoDumb} we conclude
\Begeqa
&&
\partial_\alpha \left( \frac{\delta H}{\delta \rho\dmb} \right)
\\
\nonumber
& = &
\partial_\alpha \frac{\partial f}{\partial \rho\dmb}
- \kappa \partial_\alpha \nabla^2 \rho\dmb
+ \frac{\Gamma}{\tau \tau_q} q_\beta \partial_\alpha q_\beta ,
\Endeqa
\Begeqa
&&
\rho\dmb \partial_\alpha \left( \frac{\delta H}{\delta \rho\dmb} \right)
\\
\nonumber
& = &
\partial_\alpha p\dmb
- \kappa \rho\dmb \partial_\alpha \nabla^2 \rho\dmb
+ \frac{\delta H}{\delta q_\beta} \partial_\alpha q_\beta .
\Endeqa
We may use this to eliminate the interface term in the momentum
equation:
\Begeqa
\rho D_t V_\alpha & = & - \partial_\alpha p + \partial_\alpha p\dmb
\\
\nonumber
& - & \rho\dmb \partial_\alpha \left( \frac{\delta H}{\delta \rho\dmb} \right)
+ \frac{\delta H}{\delta q_\beta} \partial_\alpha q_\beta
+ \partial_\beta \left( \frac{\delta H}{\delta q_\beta} q_\alpha \right) .
\Endeqa
We now multiply this equation with $V_\alpha$ and integrate over
space. On the left hand side, this yields, making use of
Eq.~\ref{eq:UsefulDerivativeTransformation},
\Begeqa
\int d^3\vec{r} V_\alpha \rho D_t V_\alpha & = &
\int d^3\vec{r} \rho V_\alpha \partial_t V_\alpha
\\
\nonumber
& = &
\int d^3\vec{r} \frac{\delta H}{\delta \vec{V}} \cdot \partial_t \vec{V} .
\Endeqa
On the right hand side, we employ a few integrations by parts and
incompressibility, and also insert the other equations of motion. In
particular, we obtain the term
\Begeqa
& &
- \int d^3\vec{r} V_\alpha \rho\dmb
\partial_\alpha \left( \frac{\delta H}{\delta \rho\dmb} \right)
\\
\nonumber
& = &
\int d^3\vec{r} \frac{\delta H}{\delta \rho\dmb}
V_\alpha \partial_\alpha \rho\dmb
=
\int d^3\vec{r} \frac{\delta H}{\delta \rho\dmb}
\left( D_t - \partial_t \right) \rho\dmb
\\
\nonumber
& = &
- \int d^3\vec{r} \frac{\delta H}{\delta \rho\dmb} \partial_t \rho\dmb 
\Endeqa
as well as
\Begeqa
& &
\int d^3\vec{r} V_\alpha
\frac{\delta H}{\delta q_\beta} \partial_\alpha q_\beta
=
\int d^3\vec{r} \frac{\delta H}{\delta q_\beta}
\left( D_t - \partial_t \right) q_\beta
\\
\nonumber
& = &
- \int d^3\vec{r} \frac{\delta H}{\delta \vec{q}}
\cdot \partial_t \vec{q}
+ \int d^3\vec{r} \frac{\delta H}{\delta q_\beta}
q_\alpha \partial_\alpha V_\beta
\\
\nonumber
& = &
- \int d^3\vec{r} \frac{\delta H}{\delta \vec{q}} \cdot \partial_t \vec{q}
- \int d^3\vec{r} V_\beta \partial_\alpha
\left( \frac{\delta H}{\delta q_\beta} q_\alpha \right)
\Endeqa
and
\Begeqa
&&
\int d^3\vec{r} V_\alpha \partial_\beta
\left( \frac{\delta H}{\delta q_\beta} q_\alpha \right)
\\
\nonumber
& = &
\int d^3\vec{r} V_\alpha \partial_\beta
\left( \frac{\delta H}{\delta q_\alpha} q_\beta \right) =
\int d^3\vec{r} V_\beta \partial_\alpha
\left( \frac{\delta H}{\delta q_\beta} q_\alpha \right) .
\Endeqa
This yields
\Begeqa
\nonumber
&&
\int d^3\vec{r} \left[
  \frac{\delta H}{\delta \vec{V}} \cdot \partial_t \vec{V} +
  \frac{\delta H}{\delta \rho\dmb} \partial_t \rho\dmb +
  \frac{\delta H}{\delta \vec{q}} \cdot \partial_t \vec{q}
  \right]
\\
& = & 0 .
\Endeqa
The left hand side of this equation amounts to $d H / dt$.

It is therefore clear that the additional terms $\vec{j}\nt$,
$\boldsymbol{\sigma}$, and $\vec{Q}$, should be of \emph{dissipative}
nature. For the dissipative part of the dynamics we thus have
\Begeqa
\label{eq:FinalApprox1Diss}
\nabla \cdot \vec{V} & = & 0 ,
\\
\label{eq:FinalApprox2Diss}
\partial_t \rho\dmb & = & - \nabla \cdot \vec{j}\nt ,
\\
\label{eq:FinalApprox3Diss}
\rho \partial_t \vec{V} & = & \eta_{\mathrm{s}} \nabla^2 \vec{V}
+ \nabla \cdot \boldsymbol{\sigma} ,
\\
\label{eq:FinalApprox4Diss}
\partial_t \vec{q} & = & - \frac{1}{\tau_q} \vec{q} + \vec{Q} .
\Endeqa
For the various contributions to the dissipation rate we obtain
\Begeqa
&&
\int d^3\vec{r} \frac{\delta H}{\delta V_\alpha} \partial_t V_\alpha
=
\int d^3\vec{r} V_\alpha \rho \partial_t V_\alpha
\\
\nonumber
& = &
\eta_{\mathrm{s}} \int d^3\vec{r} V_\alpha \partial_\beta \partial_\beta V_\alpha
+ \int d^3\vec{r} V_\alpha \partial_\beta \sigma_{\alpha \beta}
\\
\nonumber
& = &
- \eta_{\mathrm{s}} \int d^3\vec{r} (\partial_\beta V_\alpha)^2
- \int d^3\vec{r} \sigma_{\alpha \beta} \partial_\beta V_\alpha ,
\Endeqa
\Begeqa
&&
\int d^3\vec{r} \frac{\delta H}{\delta \vec{q}}
\cdot \partial_t \vec{q}
\\
\nonumber
& = &
\frac{\Gamma}{\tau \tau_q} \int d^3\vec{r} \rho\dmb
\vec{q} \cdot \left[ - \frac{1}{\tau_q} \vec{q} + \vec{Q} \right]
\\
\nonumber
& = &
- \frac{\Gamma}{\tau \tau_q^2} \int d^3\vec{r}
\rho\dmb \vec{q}^2
+ \frac{\Gamma}{\tau \tau_q} \int d^3\vec{r}
\rho\dmb \vec{q} \cdot \vec{Q} ,
\Endeqa
\Begeqa
&&
\int d^3\vec{r} \frac{\delta H}{\delta \rho\dmb} \partial_t \rho\dmb
=
- \int d^3\vec{r} \frac{\delta H}{\delta \rho\dmb}
\nabla \cdot \vec{j}\nt
\\
\nonumber
& = &
\int d^3\vec{r} \, \vec{j}\nt \cdot \nabla
\left( \frac{\delta H}{\delta \rho\dmb} \right) .
\Endeqa

In what follows, we will assume a simple model for the dissipative
terms. We will assume that $\boldsymbol{\sigma}$ and $\vec{Q}$ vanish
(or that their effect can be absorbed in a re-definition of
$\eta_{\mathrm{s}}$ and $\tau_q$), and that
\Begeq
\vec{j}\nt = - M \left( \rho\dmb \right) \nabla
\left( \frac{\delta H}{\delta \rho\dmb} \right) ,
\Endeq
where the function $M \left( \rho\dmb \right) \ge 0$ is essentially
the Onsager coefficient for interdiffusion. It is then obvious that
the relation $d H / d t \le 0$ strictly holds. It should also be noted
that, in terms of the general formalism of Section~\ref{sec:GENERIC}, we
have a matrix of dissipative terms that is diagonal and therefore
obviously symmetric. In other words, the model assumes the absence of
dissipative cross-couplings.  Whether such cross-terms are permitted
at all by symmetry, and, if yes, what form they may have, is an open
question that we do not wish to investigate here. In any case, when
introducing such terms one needs to take care that the symmetry and
the positive-definiteness of the matrix is strictly maintained. An
important observation is that the interdiffusion current is driven by
a bulk term, an interface term, and an elastic term. This last driving
force, which has apparently first been noted by Doi and
Onuki~\cite{doiDynamicCouplingStress1992}, appears here as a
straightforward consequence of the Second Law.

In summary, we have obtained the set of equations of motion
\Begeqa
\label{eq:FinalReduced1}
\nabla \cdot \vec{V} & = & 0 ,
\\
\label{eq:FinalReduced2}
D_t \rho\dmb & = & \nabla \cdot \left[ M \left( \rho\dmb \right) \nabla
\left( \frac{\delta H}{\delta \rho\dmb} \right) \right] ,
\\
\label{eq:FinalReduced3}
\rho D_t \vec{V} & = & - \nabla p
+ \eta_{\mathrm{s}} \nabla^2 \vec{V}
+ \kappa \rho\dmb \nabla \nabla^2 \rho\dmb
\\
\nonumber
&&
+ \frac{\Gamma}{\tau \tau_q}
\nabla \cdot \left( \rho\dmb \vec{q} \vec{q} \right) ,
\\
\label{eq:FinalReduced4}
D_t \vec{q} & = & \vec{q} \cdot \nabla \vec{V}
  - \frac{1}{\tau_q} \vec{q} .
\Endeqa

In principle, this concludes our derivation. To make contact with the
standard rheological literature, we transform the equations from the
vector field $\vec{q}$ to the conformation tensor field
$\boldsymbol{C} = \vec{q} \vec{q}$, which is strictly invariant under
flip $\vec{q} \to - \vec{q}$. In the momentum equation and in the
convection-diffusion equation for $\rho\dmb$, this is a simple
insertion (note $\vec{q}^2 = \mathrm{tr} \, \boldsymbol C$). The
equation of motion for $\vec{q}$ is easily transformed to
\Begeq
\label{eq:EqOfMotionConformationTensor}
D_t C_{\alpha \beta} -
C_{\alpha \gamma} \partial_\gamma V_\beta -
C_{\beta \gamma} \partial_\gamma V_\alpha
= - \frac{2}{\tau_q} C_{\alpha \beta} .
\Endeq
It should be noted that the left-hand side is nothing but the
so-called ``upper convected derivative'' known in the rheological
literature. Furthermore, it should be noted that the equation of
motion of the standard Oldroyd-B model \emph{differs} from
Eq.~\ref{eq:EqOfMotionConformationTensor}; it rather reads
\Begeqa
\label{eq:EqOfMotionConformationTensorOldroydB}
\nonumber
& & D_t C_{\alpha \beta} -
C_{\alpha \gamma} \partial_\gamma V_\beta -
C_{\beta \gamma} \partial_\gamma V_\alpha
\\
& = & - \frac{2}{\tau_q} \left[ C_{\alpha \beta}
  - \frac{k_B T}{k} \delta_{\alpha \beta} \right] ,
\Endeqa
where $k_B T$ denotes the thermal energy.
The difference can be traced back to the ensemble problems already
mentioned in the Introduction; in explicit terms this is worked out in
\ref{sec:EnsembleProblems}.

\section{Van der Waals model}
\label{sec:vanderwaals}

We have so far not yet specified what free energy we use for $f\dmb
(\rho\dmb)$. Since the solvent in our model is just a structureless
ideal gas whose main purpose is to transport momentum, we may view the
phase separation, from the point of view of thermodynamics, as just a
gas-liquid transition of the polymer component. The standard Mean
Field model for the gas-liquid transition is however the Van der Waals
model. The purpose of the present section is therefore to briefly
elucidate how the model should be modified in order to take into
account the loss of translational entropy due to chain connectivity.

\subsection{Monatomic ideal gas}

Let us first start with the well-known case of a monatomic fluid. For a
single point particle in a volume $V$, the canonical partition
function at temperature $T$ is
\Begeq
z_\mrm{id}(V,T) = \frac{V}{\Lambda^3} ,
\Endeq
where $\Lambda$ is the thermal de Brogie wavelength, which just acts
as a normalization factor to make sure $z_\mrm{id}$ is
dimensionless. For $N$ indistinguishable non-interacting particles the
partition function then is
\Begeq
  \pfc(N,V,T) = \frac{z_\mrm{id}^N}{N!}.
\Endeq
With Stirling's approximation, the corresponding Helmholtz free energy of
the ideal gas is
\Begeqa
&&
\beta F_\mrm{id} (N,V,T) = - \ln \pfc
\\
\nonumber
& = &
- N \ln \left( \frac{V}{\Lambda^3} \right) + N \ln N - N .
\Endeqa
Here $\beta = 1 / (k_B T)$, where $k_B$ is Boltzmann's constant.
The resulting pressure is the well-known expression
\Begeq
  \beta p_\mrm{id} =
  - \beta \frac{\partial F_\mrm{id}}{\partial V} = \frac{N}{V} .
\Endeq

\subsection{Monatomic Van der Waals fluid}

The Van der Waals Mean Field approximation assumes that the
multi-particle partition function may be factorized into the
product of effective single-particle partition functions,
\Begeq
  \pfc(N,V,T) = \frac{z_\mrm{eff}^N}{N!} ,
\Endeq
where the model assumes that the effects of short-range molecular
repulsion and medium-range attraction may be taken into account by (i)
an effective atomic volume $b$, such that only the volume $V - N b$ is
available for each particle, and (ii) an effective attraction energy
per atom $U$ ($U > 0$). Therefore we get
\Begeq
z_\mrm{eff} =
\frac{1}{\Lambda^3} \left( V - N b \right) \exp \left( \beta U \right) .
\Endeq
Within the Mean Field picture, $U$ should be proportional to
the probability to find another particle in the vicinity
of a test paticle, or, in other words, proportional to
the density $N / V$. We therefore write
\Begeq
U = a \frac{N}{V}
\Endeq
with a parameter $a > 0$, such that
\Begeq
z_\mrm{eff} = z_\mrm{id} \left( 1 - \frac{N b}{V} \right)
\exp \left( \beta a \frac{N}{V} \right) .
\Endeq
This results in the Helmholtz free energy
\Begeqa
  \beta F & = & -N \ln z_\mrm{eff} + N \ln N - N
  \\
  \nonumber
  & = & \beta F_\mrm{id} -
  N \ln \left( 1 - \frac{N b}{V} \right) - \beta a \frac{N^2}{V} ,
\Endeqa
and in turn in the pressure
\Begeq
\beta p = - \beta \frac{\partial F}{\partial V} =
\frac{N}{V - Nb} - \beta a \left( \frac{N}{V} \right)^2 .
\Endeq

\subsection{Polymeric Van der Waals fluid}

We now assume that the system comprises $N$ polymer chains, each of
which is in turn composed of $M$ monomers. We therefore have to deal
with two densities, (i) the number of chains per unit volume, $N / V$,
and (ii) the number of monomers per unit volume, $M N / V$.

The polymer analog of the single free particle is the single
free random walk, where all interactions are turned off, except
the bonded interactions that keep the monomers together. The
partition function of that walk is therefore
\Begeq
z_\mrm{id}(V,T) = \frac{V}{\Lambda^3} z_\mrm{conf} ,
\Endeq
where the factor $z_\mrm{conf}$ takes into account the entropy that is
associated with the different conformations of the walk.  We do not
attempt to write down an explicit expression for $z_\mrm{conf}$ but
only note that (i) it depends exponentially on $M$, and (ii) it is
independent of both $N$ and $V$. This latter independence means that
the precise form does not matter for the further development.

In analogy to the previous subsection, we may then write down the
single-walk effective partition function for a polymeric Van der
Waals system:
\Begeq
z_\mrm{eff} = z_\mrm{id} \left( 1 - \frac{N M b}{V} \right)
\exp \left( \beta a \frac{N M}{V} M \right) ;
\Endeq
here we have replaced the density $N /V$ by the monomer density $N M /
V$, which is logical when taking into account the physical origin of
the corresponding terms. The last factor of $M$ in the Boltzmann
factor takes into account that for the statistical weight of the whole
walk we have to add up all the attractions which the monomers of the
walk experience.

Therefore we obtain for the Helmholtz free energy
\Begeqa
  \beta F & = & -N \ln z_\mrm{eff} + N \ln N - N
  \\
  \nonumber
  & = & \beta F_\mrm{id} -
  N \ln \left( 1 - \frac{N M b}{V} \right) - \beta a \frac{N^2 M^2}{V} .
\Endeqa
For calculating the equation of state, we only need to take into
account that the ideal-gas pressure of the system of walks is given by
\Begeq
\beta p_\mrm{id} = \frac{N}{V} ,
\Endeq
and hence the pressure is given by
\Begeq
\beta p = - \beta \frac{\partial F}{\partial V} =
\frac{N}{V - N M b} - \beta a \left( \frac{N M}{V} \right)^2 .
\Endeq

\section{Conclusions}
\label{sec:conclus}

This paper has presented a somewhat unconventional approach to the
derivation of rheological equations for polymer solutions, which we
nevertheless consider as quite useful. Instead of the usual two-scale
description, \ie hydrodynamics on the macroscopic scale coupled to a
Fokker-Planck equation with unconstrained averaging on the
macromolecular scale, we here treat the hydrodynamic degrees of
freedom and the macromolecular ones on the same basis. The procedure
consists of (i) the definition of a sufficiently simple molecular
system, (ii) the definition of a set of fields which can be expressed
in terms of the molecular quantities, and which is chosen sufficiently
large to facilitate a consistent description of the conservative
dynamics by a Hamiltonian (Poisson bracket) formalism, (iii) the
direct construction of the corresponding dynamic equations within the
framework of a dissipative Hamiltonian system, (iv) simplification via
adiabatic elimination of fast variables, and (v) postulating
phenomenological expressions for the unknown terms, such that the
Second Law is automatically built into the description. This approach
is in spirit somewhat similar to analogous developments in the theory
of liquid crystals~\cite{starkPoissonbracketApproachDynamics2003}. As
discussed in Section~\ref{sec:DumbbellFieldTheory}, we believe that it
is physically justified to describe the polymer conformations in terms
of a vector field of end-to-end vectors. Compared to the rheology, the
``model H'' aspects are fairly straightforward to take into account.

From our model (phantom Hookean dumbbells which interact with the
solvent via Stokes friction, and with each other via a Van der Waals
background) we find a set of relatively simple equations with a
straightforward coupling of the conformation tensor to both the
momentum conservation equation and to the interdiffusion, while the
dynamics of the conformation tensor itself is, except for convection
described by the upper convected derivative, just a simple relaxation
towards zero. We believe that a nonzero average conformation tensor
should be described by coupling the whole system consistently to
Langevin noise. This extension of the model is however left for future
work.

Furthermore, it is far from clear if the model has sufficient physical
content to reproduce the rich phenomenology that is observed
experimentally for viscoelastic phase separation. This question can
only be answered by detailed simulations, and the scan of wide ranges
of parameters, which is also deferred to future work. In this context,
we would like to emphasize that the parameter $\Gamma / (\tau \tau_q)$
may be viewed as a parameter which controls the strength of the
viscoelastic coupling: If we set that parameter to zero, we recover
the standard model H. Conversely, for large values of the parameter
one should expect a strong coupling to the macromolecular internal
degree of freedom, and therefore at least some viscoelastic effects on
the phase separation dynamics.

Additional directions for future research are: (i) rigorous
mathematical analysis of the derived model with respect to existence,
weak-strong uniqueness and stability; (ii) numerical approximation and
extensive benchmarking; (iii) generalization to non-Hookean force laws
such as, \eg the FENE potential~\cite{krogerSimpleModelsComplex2004};
and (iv) a more detailed description of polymer conformations by
higher-order Rouse modes~\cite{doi_theory_1988}. It is our strong
belief that our new approach may have an important impact to new
developments of consistent and physically well-founded rheological
models.

\ack

Stimulating discussions with J. Ravi Prakash, Masao Doi, Harald
Pleiner, Jasper Michels and Sergei Egorov are gratefully
acknowledged. Funded by the Deutsche Forschungsgemeinschaft (DFG,
German Research Foundation), Project No. 233630050-TRR 146.

\begin{appendix}

\section{Poisson brackets I: General formalism}
\label{sec:poisson_general}

For the convenience of the reader, we quickly review here the Poisson
bracket formalism of Hamiltonian dynamics. We start with a set of
generalized coordinates $\{q_i\}$ and the corresponding canonically
conjugate momenta $\{p_i\}$, which together form the phase space. The
Hamiltonian $H = H \left( \{q_i\}, \{p_i\} \right)$ is assumed to not
explicitly depend on time, and the basic Hamiltonian equations of
motion are given by
\Begeqa
\label{eq:BasicHamiltonianEquations}
\dot{q_i} & = & \frac{\partial H}{\partial p_i} , \\
\dot{p_i} & = & - \frac{\partial H}{\partial q_i} .
\Endeqa
We now consider observables $f, g, h, \ldots$, \ie functions on the
phase space, where again we assume the absence of explicit time
dependence. The Poisson bracket between $f$ and $g$ is then defined via
\Begeq
\label{eq:DefinePoissonBracket}
\mybrack{f,g} = \sum_i \left(
    \frac{\partial f}{\partial q_i}
    \frac{\partial g}{\partial p_i}
  - \frac{\partial f}{\partial p_i}
    \frac{\partial g}{\partial q_i} \right) .
\Endeq
We note that $\mybrack{f,g}$ is bilinear and antisymmetric.
Furthermore, we note the elementary brackets
\Begeq
\mybrack{q_i, p_j} = \delta_{ij}
\Endeq
and the product rule
\Begeq
\mybrack{f, g h} = \mybrack{f, g} h + g \mybrack{f, h} .
\Endeq
The equation of motion of $f$ then reads, as a direct consequence of
the basic Hamiltonian equations of motion,
\Begeq
\label{eq:BasicEquationOfMotion}
\dot{f} = \mybrack{f,H} .
\Endeq
We now assume that there is some new set of variables $\{x_i\}$, in
terms of which the Hamiltonian $H$ can be conveniently written, $H = H
\left( \{x_i\} \right)$. Importantly, we do \emph{not} assume that
these variables form pairs of canonically conjugate variables. The
equation of motion for $f$ then reads
\Begeq
\dot{f} = \sum_j \mybrack{f,x_j} \frac{\partial H}{\partial x_j} .
\Endeq
In particular,
\Begeq
\dot{x_i} = \sum_j \mybrack{x_i,x_j} \frac{\partial H}{\partial x_j} .
\Endeq
If we now assume that, in terms of Poisson brackets, the new variables
form a closed set, \ie that $\mybrack{x_i, x_j}$ may be expressed
again in terms of the $x_i$, then we get a closed system of equations
where no reference to the original canonical variables needs to be
made. This is particularly useful if the $x_i$ form a small set of
collective variables.

The generalization to field theory is obvious. We assume that the
Hamiltonian $H$ is a functional of a set of fields $\Phi_i
(\vec{r})$, $H = H \left( \{ \Phi_i \} \right)$, and we
again assume that the fields form a closed set in terms of Poisson
brackets. Then the field-theoretic equations of motion are written as
\Begeq
\dot{\Phi}_i (\vec{r}) = \sum_j \int d^3 \vec{r}'
\mybrack{ \Phi_i (\vec{r}), \Phi_j (\vec{r}') }
\frac{\delta H}{\delta \Phi_j (\vec{r'})} ,
\Endeq
where $\delta H / \delta \Phi$ denotes the functional derivative. For
hydrodynamic theories, where the continuum formulation is intended to
be a simplified and coarse-grained description of an underlying
particle system, it is useful to write the fields in terms of the
microscopic coordinates and momenta, and use these representations to
evaluate the Poisson brackets.

\section{Poisson brackets II: Euler equations}
\label{sec:poisson_euler}

Let us try to elucidate that strategy via the simple example
of the Euler equations of hydrodynamics. Starting point are
the fields mass density
\Begeq
\rho (\vec{r}) = m \sum_i \delta( \vec{r} - \vec{r}_i )
\Endeq
(we assume we have a system of particles with mass $m$ located
at positions $\vec{r}_i$), and momentum density
\Begeq
\vec{j}(\vec{r}) = \sum_i \vec{p}_i \delta( \vec{r} - \vec{r}_i )
\Endeq
with particle momenta $\vec{p}_i$. We do not consider an energy or
entropy field since we are interested in isothermal hydrodynamics,
such that the conserved Hamiltonian should be interpreted as the
Helmholtz free energy of the system. One immediately finds, for
any observable $\varphi$,
\Begeq
\mybrack{ \delta ( \vec{r} - \vec{r}_i ), \varphi} =
- \frac{\partial \varphi}{\partial \vec{p}_i} \cdot
\nabla \delta ( \vec{r} - \vec{r}_i )
\Endeq
and thus
\Begeqa
\mybrack{ \rho (\vec{r}), \rho (\vec{r}') } & = & 0 , \\
\mybrack{ \rho (\vec{r}), \vec{j} (\vec{r}') } & = &
- \rho (\vec{r}') \nabla \delta ( \vec{r} - \vec{r}')  , \\
\mybrack{ j_\alpha (\vec{r}), j_\beta (\vec{r}') } & = &
j_\beta (\vec{r}) \partial_\alpha' \delta ( \vec{r}' - \vec{r})
\\
\nonumber
&&
-
j_\alpha (\vec{r}') \partial_\beta \delta ( \vec{r} - \vec{r}') ,
\Endeqa
where Greek letters denote Cartesian indexes (with Einstein summation
convention implied), and $\partial_\alpha \equiv \partial / \partial
r_\alpha$.

The transition from the particle picture to field theory is done
by replacing $\rho$ and $\vec{j}$ by ``smeared'' continuous fields.
Here it is useful to define
\Begeq
\vec{v} (\vec{r}) = \frac{\vec{j} (\vec{r})}{\rho (\vec{r})} .
\Endeq
We then postulate a field-theoretic Hamiltonian via
\Begeq
H \left[ \rho, \vec j \right] = \int d^3 \vec{r} \,
\left[ \frac{\vec{j}^2}{2 \rho} + f (\rho) \right] ,
\Endeq
where $f$ is the free energy density. The derivatives are
given by
\Begeqa
\frac{\delta H}{\delta \rho} & = &
- \frac{1}{2} \vec{v}^2 + \frac{\partial f}{\partial \rho} , \\
\frac{\delta H}{\delta \vec{j}} & = & \vec{v} .
\Endeqa
Inserting these results into the field-theoretic Hamiltonian equations
of motion, we find, after a few lines of straightforward algebra
\Begeq
\partial_t \rho + \nabla \cdot \vec{j} = 0 ,
\Endeq
\Begeq
\partial_t j_\alpha + \partial_\beta ( \rho v_\alpha v_\beta )
= - \rho \partial_\alpha
\left( \frac{\partial f}{\partial \rho} \right) ;
\Endeq
the first equation obviously is the mass conservation equation.

According to the first law, we have for the free energy per mass
\Begeq
d \left( \frac{f}{\rho} \right) =
- s dT - p d \left( \frac{1}{\rho} \right) ,
\Endeq
where $s$ is the entropy per mass, $T$ the temperature, and $p$
the pressure. In an isothermal situation, $dT = 0$, and
\Begeq
\frac{\partial}{\partial \rho} \left( \frac{f}{\rho} \right) =
\frac{p}{\rho^2}
\Endeq
or
\Begeq
p = \rho^2 \frac{\partial}{\partial \rho} \left( \frac{f}{\rho} \right)
= \rho \frac{\partial f}{\partial \rho} - f ,
\Endeq
which implies
\Begeq
\partial_\alpha p = \rho \partial_\alpha
\left( \frac{\partial f}{\partial \rho} \right) .
\Endeq
The momentum equation therefore reads
\Begeq
\partial_t j_\alpha + \partial_\beta ( \rho v_\alpha v_\beta )
= - \partial_\alpha p ,
\Endeq
which is the well-known Euler equation.

\section{Ensemble problems in meso--macro coupling}
\label{sec:EnsembleProblems}  

The present appendix attempts to elucidate in some more detail what we
mean with our remarks about ``ensemble problems'' in the main text. We
hope that formulating these considerations in a general and abstract
language helps to clarify our point.

We assume that the macroscopic domain can be divided into small cells,
such that the mesoscale description for a particular cell comprises a
set of dynamic variables $\xi_i$, $i = 1, \ldots, n$. We may \eg
assume that the $\xi_i$ describe the conformational degrees of freedom
of a polymer chain, or similar. Importantly, we assume that the
$\xi_i$ only describe \emph{internal} degrees of freedom of the
macromolecules (such as first or higher-order Rouse modes), while the
center-of-mass coordinates of the molecules are \emph{not} included
in the set.

Furthermore, we assume, in accord with the development outlined in the
monograph by Bird et al.~\cite{birdDynamicsPolymericLiquids1987a},
that the $\xi_i$ are subject to a (known) Fokker-Planck dynamics,
which describes the time evolution of the probability density
$P(\{\xi_i\},t)$. The ingredients of this description are the
(symmetric and positive-semidefinite) diffusion tensor $D_{ij} =
D_{ij} ( \{\xi_k\} )$, the mesoscopic free energy $H_{meso}
(\{\xi_i\})$, the thermal energy $k_B T$ (or $\beta = 1 / (k_B T)$),
and an additional non-equilibrium driving force $F_j$ that is
inferred from the macroscale --- for example, we may think of the
effects of a local shear flow. The Fokker-Planck equation (FPE) is
then written as
\Begeqa
\label{eq:AppFPE_FPE}
\nonumber
\partial_t P & = & \sum_{ij}
\frac{\partial}{\partial \xi_i} D_{ij}
\left( \frac{\partial}{\partial \xi_j} +
\beta \frac{\partial H_{meso}}{\partial \xi_j}
- \beta F_j \right) P
\\
& =: &
{\cal L}_{FP} P ,
\Endeqa
which defines the Fokker-Planck operator ${\cal L}_{FP}$. In the
absence of external driving ($F_j = 0$), the Boltzmann distribution $P
\propto \exp( - \beta H_{meso} )$ is an obvious stationary solution.

The meso-macro coupling is then facilitated by a set of observables
$A_i (\{\xi_j\})$, $i = 1, \ldots, m$, which are taken as additional
dynamic variables in the macroscopic equations of
motion. ``Additional'' here means ``in addition to the standard
hydrodynamic variables'' like mass and momentum. As the $A_i$ appear
at the macro--level, they should be considered as slow variables, \ie
ideally have a significantly slower dynamics than the remaining
mesoscopic degrees of freedom.

For each of the $A_i$, an additional equation of motion on the
macro-level is needed. The most reasonable dynamics that we may
assume for the $A_i$ on the macroscale is the time evolution of the
\emph{thermal averages}
\Begeq
\label{eq:AppFPE_DefineA}
A_i^{(mac)} := \left< A_i \right> = \int d^n \xi \, A_i \, P ,
\Endeq
which can be evaluated either by analytical solution of the FPE (if
feasible) or by numerical simulation. Introducing ${\cal
  L}_{FP}^{\dag}$, the adjoint Fokker-Planck operator, we may write
\Begeqa
\label{eq:AppFPE_DynA}
\nonumber
\partial_t \left< A_i \right>
& = &
\int d^n \xi \, A_i \, {\cal L}_{FP} P =
\int d^n \xi \, \left[ {\cal L}_{FP}^{\dag} A_i \right] \, P
\\
& = &
\left< {\cal L}_{FP}^{\dag} A_i \right> ,
\Endeqa
which provides an analytical form for the equation of motion if the
solution of the FPE, plus the subsequent averaging, may be calculated
analytically. However, the conceptual framework does not depend on the
analytical solvability at all, since the averages may always be
sampled by numerical simulation.

The thus-derived equations of motion for the $\left< A_i \right>$ may
then be used as additional equations on the macroscale, however with
the following ``recipe'', which takes into account that the
center-of-mass coordinates have been omitted, such that convection
effects have to be put in ``by hand'': (i) The expression $\partial_t
\left< A_i \right>$ must be replaced by $D_t A_i^{(mac)}$, where $D_t
= \partial_t + \vec V \cdot \nabla$ is the convective derivative and
$\vec V$ denotes the macroscopic flow field. (ii) If the mesoscale
description implies the evaluation of some property of the flow field
(\eg its value, or its derivatives) at a molecular center-of-mass
coordinate, the corresponding evaluation on the macroscale must be
done at the position $\vec r$, which denotes the position of the
meso-cell in the macro-domain.

As a second step, we need to consider the important back-coupling of
the mesoscale degrees of freedom to the macroscopic hydrodynamics.
This means that an additional viscoelastic stress needs to appear in
the momentum conservation equation.

The key element is here the Kramers (or virial) expression for the
stress tensor components $\Pi_{\alpha \beta}$, which allows us to
evaluate a momentary (and local) stress from a given mesoscopic
configuration. For the time being, we do not give the explicit formula
here but just write this as a function $\tilde \Pi_{\alpha \beta}
(\{\xi_i\})$. The $\xi_i$, however, do not appear in the macroscopic
description --- here we rather consider the dynamics of
$A_i^{(mac)}$. Therefore another function $\bar \Pi_{\alpha \beta}
(\{A_i^{(mac)}\})$ is needed. The important question is hence: How
should $\bar \Pi_{\alpha \beta} (\{A_i^{(mac)}\})$ be constructed from
$\tilde \Pi_{\alpha \beta} (\{\xi_i\})$?

To tackle this question, the approach put forward in the textbook by
Bird et al.~\cite{birdDynamicsPolymericLiquids1987a} starts from a
straightforward observation: Just as the time-dependent averages
$\left< A_i \right>$, we may also evaluate
\Begeq
\label{eq:AppFPE_AvStress}
\left< \tilde \Pi_{\alpha \beta} \right> =
\int d^n\xi \, \tilde \Pi_{\alpha \beta} \, P ,
\Endeq
again, either by analytic solution of the FPE, or by numerical
simulation.

From there, the textbook proceeds as follows: First, it is observed
that, for suitably chosen observables $A_i$,
Eqs.~\ref{eq:AppFPE_DefineA} and \ref{eq:AppFPE_AvStress} happen to
result in a relation that expresses $\left< \tilde \Pi_{\alpha \beta}
\right>$ as a function of the $\left< A_i \right>$:
\Begeq
\label{eq:AppFPE_BirdRelation}
\left< \tilde \Pi_{\alpha \beta} \right> =
\Sigma_{\alpha \beta} ( \{ \left< A_i \right> \}) .
\Endeq
As a matter of fact, the approach \emph{chooses} the variables $A_i$
in such a way such that the construction of such a relation becomes
possible --- or, more precisely, essentially trivial, as the $A_i$ are
(except for trivial transformations) just chosen as the components
$\tilde \Pi_{\alpha \beta} (\{\xi_i\})$. If one then assumes that this
relation can be directly transferred to the macroscale, \ie that one
should use $\bar \Pi_{\alpha \beta} (\{A_i^{(mac)}\}) = \Sigma_{\alpha
  \beta} ( \{ A_i^{(mac)} \})$, the coupling is established and the
problem ``solved''.

However, there are two aspects of the procedure which are, in our
opinion, somewhat unsatisfactory: Firstly, the choice of the $A_i$ is
not primarily driven by the notion of ``slowness'', but rather by the
technical need to obtain a stress. Fortunately, however, for polymeric
fluids the stresses \emph{are} slow variables, such that this argument
does not count very much. The second argument, though, is much more
severe: The averaging procedure of Eq.~\ref{eq:AppFPE_AvStress}
completely disregards the \emph{ensemble-defining} property of the
$A_i$. As the $A_i$ are assumed to be ``slow'', the averaging should
only be done by integrating out the remaining ``fast'' (or
``non-$A$'') degrees of freedom. Formally this means that, after
having established the macroscopic dynamics of the $A_i$ (the
functions $A_i^{(mac)} (t)$), one should evaluate the average of any
observable $X$ via the prescription
\Begeqa
\label{eq:MicroCanonicalAverage}
& &
\left[ X \right] (t)
\\
\nonumber
& = &
\frac{
  \int d^n \xi \, P \ \left[
    \Pi_{j=1}^m \delta (A_j (\{\xi_i\}) - A_j^{(mac)} (t) ) \right]
  \, X ( \{\xi_i\} )
}{
  \int d^n \xi \, P \ \left[
    \Pi_{j=1}^m \delta (A_j (\{\xi_i\}) - A_j^{(mac)} (t) ) \right]
} ,
\Endeqa
in analogy to the microcanonical ensemble. We have deliberately
introduced a new notation for this average, to distinguish it from
$\left< \ldots \right>$, which does not have any constraining delta
functions. Obviously,
\Begeq
\label{eq:IdentityOfAverages}
\left[ A_i \right] (t) = A_i^{(mac)} (t) = \left< A_i \right> (t) ,
\Endeq
while such an identity of averages does in general \emph{not} hold for
other observables. In particular, this must typically be expected for
$X = \tilde \Pi_{\alpha \beta}$.

This, in turn, means that the averaging procedure according to
Eq.~\ref{eq:MicroCanonicalAverage} will produce a stress that
\emph{differs} from the stress that results from the simple average
according to Eq.~\ref{eq:AppFPE_AvStress}. It is clear that
Eq.~\ref{eq:MicroCanonicalAverage} results in a prescription
for the macroscopic stress that reads
\Begeq
\bar \Pi_{\alpha \beta} (\{A_i^{(mac)}\}) =
\left[ \tilde \Pi_{\alpha \beta} \right] ,
\Endeq
which \emph{differs} from the prescription of
Ref.~\cite{birdDynamicsPolymericLiquids1987a},
\Begeq
\bar \Pi_{\alpha \beta} (\{A_i^{(mac)}\}) =
\left< \tilde \Pi_{\alpha \beta} \right> .
\Endeq
We strongly believe that the constrained average $\left[ \ldots
  \right]$ is more consistent with the general principles of
statistical physics than the simple average $\left< \ldots \right>$.
One obvious advantage is that the thus-constructed stress depends on
the set of macroscopic observables $A_i^{(mac)}$ \emph{by
  construction}, \emph{regardless} of how these variables are chosen.

Now, Eq.~\ref{eq:IdentityOfAverages} tells us that the problem would
not occur if the $A_i$ were permitted to be chosen to be simply the
components of the stress tensor (or the conformation tensor, which is
essentially the same object). \emph{This is however not the case.}
The components are not independent from each other, and therefore
treating each component as an independent constraint would result in
an overconstrained system. In general, we need $m \le n$ to avoid such
an overconstrained situation.

This is seen particularly easily for a simple dumbbell, which has,
beyond the (disregarded) center-of-mass coordinates, just three
degrees of freedom ($n = 3$), which we can parameterize in terms of
the connector vector $\vec{q}$. The Kramers expression for the stress
tensor is therefore
\Begeq
\tilde \Pi_{\alpha \beta} ( \vec{q} ) =
- k \frac{\rho\dmb}{m\dmb} q_\alpha q_\beta =
- k \frac{\rho\dmb}{m\dmb} C_{\alpha \beta} ;
\Endeq
here $k$, $\rho\dmb$, $m\dmb$, and $C_{\alpha \beta}$ have the same
meaning as in the main text. The kinetic part $\propto \delta_{\alpha
  \beta}$ has been ignored, since it can, for an incompressible
system, be absorbed in a re-definition of the overall
pressure. In terms of the parameterization introduced in the
main text, we may also write this as
\Begeq
\tilde \Pi_{\alpha \beta} ( \vec{q} ) =
- \frac{\Gamma}{\tau \tau_q} \rho\dmb q_\alpha q_\beta =
- \frac{\Gamma}{\tau \tau_q} \rho\dmb C_{\alpha \beta} .
\Endeq

Obviously, there are only three independent components of
the stress tensor (or the conformation tensor). Therefore, we should
pick $\{ A_i \} = \{ q_\alpha \}$ and \emph{not} $\{ A_i \} = \{
C_{\alpha \beta} \}$. As a result of this choice ($m = n$),
Eq.~\ref{eq:MicroCanonicalAverage} reduces, in this special case, to a
fairly trivial result:
\Begeqa
\label{eq:MicroCanonicalAverageSpecial}
& &
\left[ X \right] (t)
\\
\nonumber
& = &
\frac{
  \int d^3 \vec{q} \, P (\vec{q}) \, \delta (\vec{q} - \vec{q}^{(mac)})
  \, X ( \vec{q} )
}{
  \int d^3 \vec{q} \, P (\vec{q}) \, \delta (\vec{q} - \vec{q}^{(mac)})
}
\\
\nonumber
& = &
\frac{ P (\vec{q}^{(mac)}) \, X (\vec{q}^{(mac)}) }{
   P (\vec{q}^{(mac)}) } = X (\vec{q}^{(mac)}) .
\Endeqa
Therefore, we propose in the present work to use
\Begeqa
\label{eq:PiBarAccordingToSpiller}
\nonumber
\bar \Pi_{\alpha \beta} & = &
- \frac{\Gamma}{\tau \tau_q} \rho\dmb
q_\alpha^{(mac)} q_\beta^{(mac)}
\\
& = &
- \frac{\Gamma}{\tau \tau_q} \rho\dmb
\left< q_\alpha \right> \left< q_\beta \right>
\Endeqa
instead of
\Begeq
\label{eq:PiBarAccordingToBird}
\bar \Pi_{\alpha \beta} =
- \frac{\Gamma}{\tau \tau_q} \rho\dmb
C_{\alpha \beta}^{(mac)} =
- \frac{\Gamma}{\tau \tau_q} \rho\dmb
\left< q_\alpha q_\beta \right> ,
\Endeq
which would be the prescription of
Ref.~\cite{birdDynamicsPolymericLiquids1987a}. The fact that
these expressions differ significantly is a hallmark of the
strong thermal fluctuations in polymer systems.

This does \emph{not} imply that we propose to simply disregard thermal
fluctuations. We rather believe that they should be taken into account
not by a thermal average along the lines of
Eq.~\ref{eq:PiBarAccordingToBird}, but rather by explicit Langevin
noise on the macroscopic level.

Let us work out what these considerations imply for the simple case of
a Hookean dumbbell in a flow field. Again we assume a frictional
coupling of the beads to the flow field, with a friction constant
$\zeta$, and assume that second and higher-order derivatives
of the flow field may be neglected, just as in the main text.
The Fokker-Planck operator for the overdamped Brownian motion
of the dumbbell can then be constructed easily:
\Begeqa
\nonumber
{\cal L}_{FP}
& = &
- \frac{\partial}{\partial \vec{q}} \cdot
\left( - \frac{2 k}{\zeta} \vec{q}
+ \vec{q} \cdot \nabla \vec{V} \right)
\\
& + &
\frac{2 k_B T}{\zeta}
\frac{\partial^2}{\partial \vec{q}^2} ;
\Endeqa
here $\nabla \vec{V}$ is the gradient of the velocity field at the
position of the dumbbell's center of mass. In terms of the
parameterization of the main text we may write this as
\Begeqa
\nonumber
{\cal L}_{FP}
& = &
- \frac{\partial}{\partial \vec{q}} \cdot
\left( - \frac{1}{\tau_q} \vec{q}
+ \vec{q} \cdot \nabla \vec{V} \right)
\\
& + &
\frac{k_B T}{k \tau_q}
\frac{\partial^2}{\partial \vec{q}^2} .
\Endeqa
The adjoint operator is then found as
\Begeqa
\nonumber
{\cal L}_{FP}^{\dag}
& = &
\left( - \frac{1}{\tau_q} \vec{q}
+ \vec{q} \cdot \nabla \vec{V} \right)
\cdot \frac{\partial}{\partial \vec{q}}
\\
& + &
\frac{k_B T}{k \tau_q}
\frac{\partial^2}{\partial \vec{q}^2} ,
\Endeqa
from which we evaluate
\Begeq
{\cal L}_{FP}^{\dag} \vec{q} =
- \frac{1}{\tau_q} \vec{q}
+ \vec{q} \cdot \nabla \vec{V}
\Endeq
and
\Begeqa
& &
{\cal L}_{FP}^{\dag} q_\alpha q_\beta
\\
\nonumber
& = &
q_\alpha q_\gamma \partial_\gamma V_\beta +
q_\beta q_\gamma \partial_\gamma V_\alpha -
\frac{2}{\tau_q} q_\alpha q_\beta +
\frac{2 k_B T}{k \tau_q} \delta_{\alpha \beta} ,
\Endeqa
which in turn implies
\Begeq
\label{eq:FPEApp1stMoment}
\partial_t \left< \vec{q} \right>
= \left< {\cal L}_{FP}^{\dag} \vec{q} \right>
= 
- \frac{1}{\tau_q} \left< \vec{q} \right>
+ \left< \vec{q} \right> \cdot \nabla \vec{V}
\Endeq
and
\Begeqa
\label{eq:FPEApp2ndMoment}
& &
\partial_t \left< q_\alpha q_\beta \right>
= \left< {\cal L}_{FP}^{\dag} q_\alpha q_\beta \right>
\\
\nonumber
& = &
\left< q_\alpha q_\gamma \right> \partial_\gamma V_\beta +
\left< q_\beta q_\gamma \right> \partial_\gamma V_\alpha -
\frac{2}{\tau_q} \left< q_\alpha q_\beta \right> +
\frac{2 k_B T}{k \tau_q} \delta_{\alpha \beta} .
\Endeqa
From Eq.~\ref{eq:FPEApp1stMoment} we find
\Begeqa
& &
\partial_t \left( \left< q_\alpha \right>
\left< q_\beta \right> \right)
\\
\nonumber
& = &
\left< q_\alpha \right> \left< q_\gamma \right> \partial_\gamma V_\beta +
\left< q_\beta \right> \left< q_\gamma \right> \partial_\gamma V_\alpha -
\frac{2}{\tau_q} \left< q_\alpha \right> \left< q_\beta \right> .
\Endeqa
After replacing the partial time derivative with the convective
derivative, we see that this is precisely the equation of motion for
the conformation tensor that has been derived in the main text. In
other words, the present paper proposes to define the macroscopic
conformation tensor as $\left< q_\alpha \right> \left< q_\beta
\right>$. Conversely, Ref.~\cite{birdDynamicsPolymericLiquids1987a}
(or, in other words, the standard Oldroyd-B model) proposes to use
$\left< q_\alpha q_\beta \right>$, whose equation of motion differs
from the one proposed here by the term $[(2 k_B T) / (k \tau_q)]
\delta_{\alpha \beta}$. The fact that the difference is propotional to
the thermal energy $k_B T$ makes it obvious that the difference lies
in the different treatment of thermal fluctuations.

In equilibrium, where $\partial_t (\ldots) = 0$, $\vec{V} = 0$, we
have $\left< q_\alpha \right> \left< q_\beta \right> = 0$, while
$\left< q_\alpha q_\beta \right> = [(k_B T) / k] \delta_{\alpha
  \beta}$ in accord with the equipartition theorem. We may therefore
say that the prescription of the present paper implies a relaxation
towards \emph{mechanical} equilibrium, while the prescription of
Ref.~\cite{birdDynamicsPolymericLiquids1987a} implies relaxation
towards \emph{thermal} equilibrium.

\end{appendix}

\onecolumn

\section*{References}

\begin{flushleft}
%
% \bibliographystyle{iopart-num}
% \bibliography{EquViscPhaseSepPaper}

\begin{thebibliography}{10}
\expandafter\ifx\csname url\endcsname\relax
  \def\url#1{{\tt #1}}\fi
\expandafter\ifx\csname urlprefix\endcsname\relax\def\urlprefix{URL }\fi
\providecommand{\eprint}[2][]{\url{#2}}

\bibitem{onukiPhaseTransitionDynamics2002a}
Onuki A 2002 {\em Phase {Transition} {Dynamics}\/} (Cambridge University Press)
  ISBN 978-1-139-43316-7

\bibitem{brayTheoryPhaseorderingKinetics2002}
Bray A~J 2002 {\em Advances in Physics\/} {\bf 51} 481--587 ISSN 0001-8732
  \urlprefix\url{http://dx.doi.org/10.1080/00018730110117433}

\bibitem{chaikinPrinciplesCondensedMatter2000}
Chaikin P~M and Lubensky T~C 2000 {\em Principles of {Condensed} {Matter}
  {Physics}\/} (Cambridge University Press) ISBN 978-0-521-79450-3

\bibitem{tanakaUnusualPhaseSeparation1993}
Tanaka H 1993 {\em Physical Review Letters\/} {\bf 71} 3158--3161 publisher:
  American Physical Society
  \urlprefix\url{https://link.aps.org/doi/10.1103/PhysRevLett.71.3158}

\bibitem{tanakaUniversalityViscoelasticPhase1996}
Tanaka H 1996 {\em Physical Review Letters\/} {\bf 76} 787--790 publisher:
  American Physical Society
  \urlprefix\url{https://link.aps.org/doi/10.1103/PhysRevLett.76.787}

\bibitem{tanakaViscoelasticModelPhase1997}
Tanaka H 1997 {\em Physical Review E\/} {\bf 56} 4451--4462 publisher: American
  Physical Society
  \urlprefix\url{https://link.aps.org/doi/10.1103/PhysRevE.56.4451}

\bibitem{tanakaRolesBulkRelaxation1997}
Tanaka H 1997 {\em Progress of Theoretical Physics Supplement\/} {\bf 126}
  333--338 ISSN 0375-9687 publisher: Oxford Academic
  \urlprefix\url{https://academic.oup.com/ptps/article/doi/10.1143/PTP.126.333/1944573}

\bibitem{tanakaViscoelasticModelPhase1999}
Tanaka H 1999 {\em Physical Review E\/} {\bf 59} 6842--6852 publisher: American
  Physical Society
  \urlprefix\url{https://link.aps.org/doi/10.1103/PhysRevE.59.6842}

\bibitem{tanakaViscoelasticModelPhase2000}
Tanaka H 2000 {\em AIP Conference Proceedings\/} {\bf 519} 52--63 ISSN
  0094-243X \urlprefix\url{http://aip.scitation.org/doi/abs/10.1063/1.1291521}

\bibitem{tanakaViscoelasticPhaseSeparation2000a}
Tanaka H 2000 {\em Journal of Physics: Condensed Matter\/} {\bf 12} R207--R264
  ISSN 0953-8984

\bibitem{nakazawaPhaseSeparationGelation2001}
Nakazawa H, Fujinami S, Motoyama M, Ohta T, Araki T, Tanaka H, Fujisawa T,
  Nakada H, Hayashi M and Aizawa M 2001 {\em Computational and Theoretical
  Polymer Science\/} {\bf 11} 445--458 ISSN 1089-3156
  \urlprefix\url{http://www.sciencedirect.com/science/article/pii/S1089315601000307}

\bibitem{tanakaNetworkFormationViscoelastic2002}
Tanaka H, Koyama T and Araki T 2002 {\em Journal of Physics: Condensed
  Matter\/} {\bf 15} S263--S268 ISSN 0953-8984 publisher: IOP Publishing

\bibitem{tanakaUniversalityViscoelasticPhase2005}
Tanaka H, Araki T, Koyama T and Nishikawa Y 2005 {\em Journal of Physics:
  Condensed Matter\/} {\bf 17} S3195--S3204 ISSN 0953-8984 publisher: IOP
  Publishing

\bibitem{tatenoPowerlawCoarseningNetworkforming2021}
Tateno M and Tanaka H 2021 {\em Nature Communications\/} {\bf 12} 912 ISSN
  2041-1723 \urlprefix\url{https://www.nature.com/articles/s41467-020-20734-8}

\bibitem{birdDynamicsPolymericLiquids1987}
Bird R~B, Armstrong R~C and Hassager O 1987 {\em Dynamics of {Polymeric}
  {Liquids}, {Volume} 1: {Fluid} {Mechanics}\/} (Wiley) ISBN 978-0-471-80245-7

\bibitem{birdDynamicsPolymericLiquids1987a}
Bird R~B, Curtiss C~F, Armstrong R~C and Hassager O 1987 {\em Dynamics of
  {Polymeric} {Liquids}, {Volume} 2: {Kinetic} {Theory}\/} (Wiley) ISBN
  978-0-471-80244-0

\bibitem{taniguchiNetworkDomainStructure1996}
Taniguchi T and Onuki A 1996 {\em Physical Review Letters\/} {\bf 77}
  4910--4913 publisher: American Physical Society
  \urlprefix\url{https://link.aps.org/doi/10.1103/PhysRevLett.77.4910}

\bibitem{zhouModifiedModelsPolymer2006a}
Zhou D, Zhang P and Weinan E 2006 {\em Physical Review E\/} {\bf 73} 061801

\bibitem{doiDynamicCouplingStress1992}
Doi M and Onuki A 1992 {\em Journal de Physique II\/} {\bf 2} 1631--1656 ISSN
  1155-4312, 1286-4870
  \urlprefix\url{http://www.edpsciences.org/10.1051/jp2:1992225}

\bibitem{milnerDynamicalTheoryConcentration1993}
Milner S~T 1993 {\em Physical Review E\/} {\bf 48} 3674--3691 publisher:
  American Physical Society
  \urlprefix\url{https://link.aps.org/doi/10.1103/PhysRevE.48.3674}

\bibitem{elafifRheologyDiffusionSimple1999}
Elafif A, Grmela M and Lebon G 1999 {\em Journal of Non-Newtonian Fluid
  Mechanics\/} {\bf 86} 253--275 ISSN 0377-0257
  \urlprefix\url{http://www.sciencedirect.com/science/article/pii/S0377025798002110}

\bibitem{pleinerGeneralNonlinear2Fluid2004}
Pleiner H and Harden J~L 2004 {\em AIP Conference Proceedings\/} {\bf 708}
  46--51 ISSN 0094-243X
  \urlprefix\url{http://aip.scitation.org/doi/abs/10.1063/1.1764058}

\bibitem{starkPoissonbracketApproachDynamics2003}
Stark H and Lubensky T~C 2003 {\em Physical Review E\/} {\bf 67} 061709
  publisher: American Physical Society
  \urlprefix\url{https://link.aps.org/doi/10.1103/PhysRevE.67.061709}

\bibitem{salmonHamiltonianFluidMechanics1988}
Salmon R 1988 {\em Annual Review of Fluid Mechanics\/} {\bf 20} 225--256 ISSN
  0066-4189 publisher: Annual Reviews
  \urlprefix\url{https://www.annualreviews.org/doi/10.1146/annurev.fl.20.010188.001301}

\bibitem{zakharovHamiltonianFormalismNonlinear1997a}
Zakharov V~E and Kuznetsov E~A 1997 {\em Physics-Uspekhi\/} {\bf 40} 1087 ISSN
  1063-7869
  \urlprefix\url{http://iopscience.iop.org/article/10.1070/PU1997v040n11ABEH000304/meta}

\bibitem{morrisonHamiltonianDescriptionIdeal1998}
Morrison P~J 1998 {\em Reviews of Modern Physics\/} {\bf 70} 467--521
  publisher: American Physical Society
  \urlprefix\url{https://link.aps.org/doi/10.1103/RevModPhys.70.467}

\bibitem{berisPoissonBracketFormulation1990}
Beris A~N and Edwards B~J 1990 {\em Journal of Rheology\/} {\bf 34} 55--78 ISSN
  0148-6055 publisher: The Society of Rheology
  \urlprefix\url{http://sor.scitation.org/doi/abs/10.1122/1.550114}

\bibitem{berisPoissonBracketFormulation1990a}
Beris A~N and Edwards B~J 1990 {\em Journal of Rheology\/} {\bf 34} 503--538
  ISSN 0148-6055 publisher: The Society of Rheology
  \urlprefix\url{http://sor.scitation.org/doi/abs/10.1122/1.550094}

\bibitem{edwardsNoncanonicalPoissonBracket1991}
Edwards B~J and Beris A~N 1991 {\em Journal of Physics A: Mathematical and
  General\/} {\bf 24} 2461--2480 ISSN 0305-4470

\bibitem{berisThermodynamicsFlowingSystems1994}
Beris A~N and Edwards B~J 1994 {\em Thermodynamics of {Flowing} {Systems}:
  {With} {Internal} {Microstructure}\/} (Oxford University Press) ISBN
  978-0-19-507694-3

\bibitem{ottingerEquilibriumThermodynamics2005}
Öttinger H~C 2005 {\em Beyond {Equilibrium} {Thermodynamics}\/} (John Wiley \&
  Sons) ISBN 978-0-471-72791-0

\bibitem{grmelaDynamicsThermodynamicsComplex1997}
Grmela M and \"{O}ttinger H~C 1997 {\em Physical Review E\/} {\bf 56}
  6620--6632 publisher: American Physical Society
  \urlprefix\url{https://link.aps.org/doi/10.1103/PhysRevE.56.6620}

\bibitem{ottingerDynamicsThermodynamicsComplex1997}
\"{O}ttinger H~C and Grmela M 1997 {\em Physical Review E\/} {\bf 56}
  6633--6655 publisher: American Physical Society
  \urlprefix\url{https://link.aps.org/doi/10.1103/PhysRevE.56.6633}

\bibitem{grmelaWhyGENERIC2010}
Grmela M 2010 {\em Journal of Non-Newtonian Fluid Mechanics\/} {\bf 165}
  980--986 ISSN 0377-0257
  \urlprefix\url{http://www.sciencedirect.com/science/article/pii/S0377025710000200}

\bibitem{lukacova-medvidovaEnergyStableNumerical2016}
\MariasLastName\ M, D\"{u}nweg B, Strasser P and Tretyakov N 2016
  {Energy}-{Stable} {Numerical} {Schemes} for {Multiscale} {Simulations} of
  {Polymer}–{Solvent} {Mixtures} {\em Mathematical {Analysis} of {Continuum}
  {Mechanics} and {Industrial} {Applications} {II}\/} Mathematics for
  {Industry} (Springer, Singapore) pp 153--165 ISBN 978-981-10-6282-7
  978-981-10-6283-4

\bibitem{strasserEnergystableLinear2019}
Strasser P~J, Tierra G, D\"{u}nweg B and \MariasLastName\ M 2019 {\em
  Computers \& Mathematics with Applications\/} {\bf 77} 125--143 ISSN
  0898-1221
  \urlprefix\url{http://www.sciencedirect.com/science/article/pii/S0898122118305303}

\bibitem{dunwegLatticeBoltzmann2009}
D\"{u}nweg B and Ladd A~J~C 2009 {Lattice} {Boltzmann} {Simulations} of {Soft}
  {Matter} {Systems} {\em Advanced {Computer} {Simulation} {Approaches} for
  {Soft} {Matter} {Sciences} {III}\/} ({\em Advances in {Polymer} {Science}\/}
  no 221) ed Holm C and Kremer K (Springer Berlin Heidelberg) pp 89--166 ISBN
  978-3-540-87705-9 978-3-540-87706-6

\bibitem{ahlrichsSimulationSingle1999}
Ahlrichs P and D\"{u}nweg B 1999 {\em The Journal of Chemical Physics\/} {\bf
  111} 8225--8239 ISSN 0021-9606
  \urlprefix\url{http://aip.scitation.org/doi/abs/10.1063/1.480156}

\bibitem{tretyakovImprovedDissipative2017}
Tretyakov N and D\"{u}nweg B 2017 {\em Computer Physics Communications\/} {\bf
  216} 102--108 ISSN 0010-4655
  \urlprefix\url{http://www.sciencedirect.com/science/article/pii/S0010465517300966}

\bibitem{langerIntroductionKineticsFirstorder1992}
Langer J~S 1992 An introduction to the kinetics of first-order phase
  transitions {\em Solids far from equilibrium\/} Collection {Alea}-{Saclay}
  {Monographs} and {Texts} in {Statistical} {Physics} ed Godreche C (Cambridge:
  Cambridge University Press) pp 297--363 ISBN 0-521-41170-X

\bibitem{ivanchenkoPhysicsCriticalFluctuations1995}
Ivanchenko Y~M and Lisyansky A~A 1995 {\em Physics of {Critical}
  {Fluctuations}\/} Graduate {Texts} in {Contemporary} {Physics} (New York, NY:
  Springer) ISBN 978-1-4612-4204-8
  \urlprefix\url{https://link.springer.com/book/10.1007/978-1-4612-4204-8}

\bibitem{doi_theory_1988}
Doi M and Edwards S~F 1988 {\em The theory of polymer dynamics\/} (Oxford
  University Press) ISBN 978-0-19-852033-7

\bibitem{krogerSimpleModelsComplex2004}
Kröger M 2004 {\em Physics Reports\/} {\bf 390} 453--551 ISSN 0370-1573
  \urlprefix\url{http://www.sciencedirect.com/science/article/pii/S0370157303003958}

\end{thebibliography}
%
\providecommand{\newblock}{}

\end{flushleft}

\end{document}